\documentclass[aps,prd,twocolumn,superscriptaddress,preprintnumbers,floatfix]{revtex4-1}
\usepackage{amsmath}
\usepackage{mathrsfs}
\usepackage{slashed}
\usepackage{gensymb}
\usepackage{amsfonts}
\usepackage{mathrsfs}
\usepackage{bbding}
\usepackage{graphicx}
\usepackage{color}
\usepackage{xcolor}
\usepackage{siunitx}
\usepackage{float}
\usepackage[colorlinks=true, linkcolor=blue, citecolor=green]{hyperref}
\usepackage{subfigure}
\usepackage[utf8]{inputenc}

\begin{document}
\title{Parameter Estimation of the Gravitational-Wave Angular Power Spectrum in the Dirty-Map Space}
\author{Erik T. Floden}
\affiliation{School of Physics and Astronomy, University of Minnesota, Minneapolis, MN 55455, USA}
\author{Alex E. Granados}
\affiliation{School of Physics and Astronomy, University of Minnesota, Minneapolis, MN 55455, USA}
\author{Vuk Mandic}
\affiliation{School of Physics and Astronomy, University of Minnesota, Minneapolis, MN 55455, USA}




\date{\today}
\begin{abstract}
We consider a search for the anisotropic stochastic gravitational-wave background (SGWB) that decomposes the sky map into its spherical harmonics components in order to obtain estimators of the angular power spectrum. Such a search often requires the inversion of a Fisher information matrix which contains small singular values. Rather than dealing with biases induced by regularization methods used to facilitate this matrix inversion, we opt to avoid this inversion step entirely by working in the so-called ``dirty map" space, and we introduce methodology for statistical model inference  in this space. We apply our methodology to simulated model signals added to detector noise characterized by Advanced LIGO's third observing run and consider angular power spectra for both the SGWB auto-correlation search as well as a cross-correlation search between the SGWB and electromagnetic tracers of matter structure in the universe. In both cases we are able to reliably recover simulated model parameters for sufficiently strong signals up to maximum order spherical harmonic modes of $\ell_{max}=10$. We find the limitations of our methodology to arise from the computational cost of testing complex models, the assumption of Gaussianity of the angular power spectrum, and a cosmic variance-like source of uncertainty which scales with the strength of the underlying signal. 
\end{abstract}
\maketitle
\section{Introduction}\label{sec:intro}

A decade ago, the direct observation of gravitational waves (GWs) opened a new window into the Universe, enabling tests of general relativity and novel probes of astrophysical populations \cite{gw150914}.
The Laser Interferometer Gravitational-wave Observatory (LIGO), located in Hanford, Washington, and Livingston, Louisiana, together with the Virgo detector in Pisa, Italy and KAGRA in Hida, Japan, have been at the forefront of these discoveries \cite{ALIGO, AdvancedVirgo,kagra}.
These interferometers have completed their fourth observing run (O4) and to date have detected over 200 Compact Binary Coalescence (CBC) candidates, hence enabling broad GW astrophysics research \cite{GWTC4}.

A target of current searches is the Stochastic Gravitational Wave Background (SGWB), the superposition of unresolved GW sources~\cite{maggiore}.
The SGWB arises from a number of astrophysical sources (e.g. unresolved CBCs) \cite{WuEA_2012, ZhuEA_2013, Ferrari_ccs, ZhuEA_2010, ZhuEA_2011_rm, Lasky_2015} and cosmological sources (e.g. early universe inflation)\cite{1976JPhA....9.1387K, SiemensEA_2007, Marzola:2017jzl, vonHarling:2019gme, starobinksii, Turner_1997}.
Evidence for a stochastic signal at nanohertz frequencies has recently been reported by pulsar timing arrays, possibly corresponding to super-massive black hole mergers and highlighting the promise of stochastic searches across multiple frequency bands \cite{Agazie_2023, ipta2023comparing}.
In the LIGO–Virgo-KAGRA (LVK) band, a detection of the SGWB would provide a complementary perspective on GW astrophysics and cosmology.

While we expect the SGWB to be isotropic to first order, astrophysical and cosmological phenomena are also expected to give rise to anisotropies in the SGWB \cite{Contaldi_2017, jenkins2019, Jenkins_2019_2, Pitrou_2020}. These anisotropies may be used to determine distributions of GW sources within our galaxy as well as the large-scale structure of the universe \cite{Cusin:2017fwz, Cusin:2017mjm, Cusin_2018_2, Cusin:2019jpv}. Anisotropies in the distribution of GWs resulting from topological phase transitions may provide information on inhomogeneities in the early universe \cite{Geller}, and the motion of terrestrial observers relative to the GW rest frame can give rise to kinematic anisotropies \cite{Allen:1996gp,Cusin:2022cbb,ValbusaDallArmi:2022htu,Chung:2022xhv}. Additionally, the propagation of GWs through inhomogeneities of the universe's mass distribution can also contribute to the background's anisotropy \cite{Bertacca_2020}. 

GW time-series strain data from multiple spatially-separated detectors can be cross-correlated to search for the anisotropic SGWB. One type of search involves decomposing the map of the GW sky into the spherical harmonics basis and estimating the corresponding coefficients of these basis functions. From these coefficients one can construct the SGWB angular power spectrum, which measures the amount of GW power distributed over different angular scales on the sky. While analyses of Advanced LIGO and Advanced Virgo's first four observing runs have not yet yielded a detection of the anisotropic SGWB, upper limits have been placed on the values of the angular power spectrum $A_\ell$ for various spectral models, including that of an astrophysical background of CBC events, which will be the main focus of this paper. These upper limit calculations were performed for maximum spherical harmonic multipoles of $\ell_{max}=4$ and resulted in limits $A_\ell < 10^{-20}$ sr$^{-2}$ \cite{o1_directional, Abbott:O2_aniso, o3aniso}.

Some SGWB searches correlate the SGWB with electromagnetic (EM) tracers of structures in the universe, such as galaxy counts, weak lensing, or the Cosmic Microwave Background (CMB) \cite{Yang_2020,Yang_2023, Cusin_2017, Cusin_2019}.
The EM tracers are expected to be correlated with different astrophysical and cosmological models that give rise to the SGWB \cite{Yang_2020,Yang_2023, Cusin_2017, Cusin_2019}.
Hence, this approach simultaneously probes the sources of SGWB anisotropy and has the potential to aid in an indirect detection of the anisotropic SGWB sooner than direct detection methods \cite{Yang_2020,Yang_2023, Cusin_2017, Cusin_2019}.
Similarly to the SGWB-only analysis, SGWB-EM correlation measurements also focus on the angular power spectrum, but the spectrum is constructed from the cross-correlation between the SGWB and EM tracer sky distribution maps.
This analysis approach is explained in greater detail in Sec \ref{sec:GWEM-clean}.
Cross-correlation studies between the SGWB and galaxy count (GC) distribution using LIGO O3 data and Sloan Digital Sky Survey data resulted in upper limits on the cross-correlation angular power spectrum on the order of $10^{-14}$, with values varying slightly depending on the angular scale up to $\ell_{max}=5$~\cite{Yang_2020,Yang_2023}. These studies also investigated implications of their measurements for astrophysical SGWB models due to binary black hole mergers, and they placed first constraints on such models  
\cite{Yang_2023}.

Both the SGWB auto-power case and the SGWB-EM cross-power case are limited in their angular resolution. As already stated, the auto-power case has a limit of $\ell_{max}=4$ and the cross-power case has a limit of $\ell_{max}=5$ \cite{Yang_2023}. These limitations come from the sensitivity of the detector network, which is captured in the co-called {\it Fisher matrix}, defined in Section II. Fisher matrix depends both on the detectors' strain sensitivities and on the geometry of the detectors' response to SGWB anisotropy. Deconvolving the detectors response from the SGWB measurement requires inversion of the Fisher matrix, as we will see below. 
However, a given network of GW detectors may be insensitive to certain SGWB sky modes, which results in an ill-defined Fisher matrix, making inversion difficult \cite{sph_methods}.
Current methods of inversion involve regularizing the Fisher matrix through variations of the Single Value Decomposition (SVD) technique \cite{sph_methods,o3aniso,Yang_2023}.
These regularization schemes can suppress or bias features in the recovered SGWB sky maps, implying limitations to the angular resolution \cite{sph_methods,o3aniso,Yang_2023}.

We propose a statistical inference method using SGWB auto-correlation and SGWB-EM cross-correlation spectra that avoids inverting an ill-defined Fisher matrix, thereby increasing the angular resolution achievable by the search. In Section II we review the standard inference formalisms, and in Section III we introduce the new formalisms of our approach. In Section IV we present results of our method applied to simulated data, and we conclude in Section V.

\section{Inference in the Clean Space}\label{sec:Cl}

\subsection{SGWB Angular Power Spectrum}
When performing an auto-power search for the SGWB using the spherical harmonics decomposition of the sky, the resulting map of GW power on the sky is given by \cite{sph_methods} 
\begin{align}\label{eq:dirty}
		\hat{x}_{\ell m}=\sum_{f,t}\gamma_{\ell m}^* (f,t)\frac{H(f)}{P_1(f,t) P_2(f,t)}\hat{C}(f,t).
	\end{align}
Here, $H(f)$ is the frequency power spectrum of the SGWB model, taken to be a power-law,
     \begin{align}
     H(f) =\left(\frac{f}{f_{\rm ref}}\right)^{\alpha -3},
     \end{align}
where the spectral index $\alpha$ takes on different values for different SGWB models and $f_{ref}$ is some reference frequency, usually chosen to be 25 Hz \cite{o3aniso}. Eq. \ref{eq:dirty} also includes the one-sided power spectra of two detectors $P_1(f,t),P_2(f,t)$, the overlap-reduction factor $\gamma_{\ell m}(f,t)$, and the cross-correlation spectra $\hat{C}(f,t)$. In these quantities we use $t$ to denote time segments into which the data is divided, and we refer the reader to \cite{sph_methods} for more details regarding these quantities. The map defined in Eq. \ref{eq:dirty} represents the GW power convolved with the response of the detectors. For this reason this map is referred to as the ``dirty" map. The covariance matrix for the elements of the dirty map is referred to as the Fisher matrix and given as
	\begin{align}\label{eq:fisher}
		\Gamma_{\ell m, \ell' m'} = \sum_{f,t} \gamma_{\ell m}^*(f,t) \frac{H^2(f)}{P_1(f,t) P_2(f,t)}\gamma_{\ell' m'}(f,t).
	\end{align}
Previous searches for the SGWB in the spherical harmonics basis  deconvolve the GW signal from the detector response. The resulting estimators of the elements of the ``clean" map are given by 
	\begin{align}\label{eq:p0}
		\hat{a}_{\ell m}=\sum_{\ell' m'} \big(\Gamma_{\rm R}^{-1}\big)_{\ell m, \ell' m'}\hat{x}_{\ell' m'}.
	\end{align}

Note that obtaining estimators of the clean map requires inversion of the Fisher matrix. Gaps in the sensitivity of the detector network manifest as small singular values in the Fisher matrix. These small singular values of $\Gamma$ result in large uncertainties in the estimators of the clean map elements. Regularization of $\Gamma$ is often performed through the use of singular value decomposition (SVD) to reduce the uncertainty in the clean map estimators. The regularization method employed by the most recent searches involves reassigning the smallest one-third of the eigenvalues of $\Gamma$ to infinity \cite{o1_directional,Abbott:O2_aniso, o3aniso}. The subscript $\rm R$ in Eq. \ref{eq:p0} denotes that regularization was performed in calculating $\Gamma^{-1}$. 

Finally the estimators of the clean map are used to calculate estimators of the angular power spectrum:
	\begin{align}\label{eq:A_ell}
		\hat{A_{\ell}}=\left(\frac{2\pi^2f_{\rm ref}^3}{3H_0^2}\right)^2\frac{1}{1+2\ell}\sum_{m=-\ell}^{\ell} \left[|\hat{a}_{\ell m}|^2 - (\Gamma^{-1}_R)_{\ell m,\ell m}\right],
	\end{align}
where $H_0$ is the Hubble constant. The second term in the sum in Eq. \ref{eq:A_ell} is a bias correction term which comes from the definitions of the expectation value and variance (see Appendix \ref{ap:estimator_auto}):
    \begin{align}\label{eq:Exp1}
    \langle \hat{x}_{\ell m}\hat{x}^*_{\ell'm'}\rangle - \langle \hat{x}_{\ell m}\rangle \langle \hat{x}^*_{\ell'm'}\rangle \approx \Gamma_{\ell m \ell' m'}
    \\
    \langle \hat{a}_{\ell m}\hat{a}^*_{\ell'm'}\rangle - \langle \hat{a}_{\ell m}\rangle \langle \hat{a}^*_{\ell'm'}\rangle \approx \Gamma^{-1}_{\ell m \ell' m'}.
    \end{align}
    
Since elements of the dirty map result from summing over many time segments of data, they are well-approximated as Gaussian-distributed random variables. The corresponding elements of the clean map are therefore also Gaussian-distributed. Note, therefore, that the angular power spectrum is a sum of squared Gaussian random variables and therefore not Gaussian. When $\ell$ is large, the sum in Eq. \ref{eq:A_ell} is performed over enough $\ell m$ modes that the estimators $\hat{A_\ell}$ may be approximated as Gaussian, but in the case of the SGWB search, the maximum spherical harmonics mode $\ell_{max}=4$ is typically used for a background of compact binary coalescences. 

With this limitation in mind, in this study we will approximate the angular power spectrum as Gaussian-distributed when developing a Bayesian likelihood analysis. Specifically, we use a Gaussian likelihood to compare our estimators of the angular power spectrum with model predictions $A_l^M(\boldsymbol{\theta})$ parameterized by a set of parameters $\boldsymbol{\theta}$:

\begin{widetext}
    \begin{equation}\label{eqn:lnLike_clean_auto}
        \ln{\mathcal{L}}\left( \hat{A}_\ell \mid A_\ell^{M}(\boldsymbol{\theta}) \right) \propto -\frac{1}{2} \left( \hat{A}_\ell - A_\ell^{M}(\boldsymbol{\theta}) \right) (K^A)^{-1}_{\ell,\ell} \left( \hat{A}_\ell - A_\ell^{M}(\boldsymbol{\theta}) \right)
    \end{equation}
\end{widetext}

where the covariance matrix $K^A$ is given by \cite{zhang_2025}:
\begin{widetext}

\begin{equation}\label{eqn:KFisherClean_auto}
    K^A_{\ell, \ell '}=\left( \frac{2\pi^2f^3}{3H_0^2}\right)^4\sum_{m ,m'} \frac{  |( {\Gamma}_R^{-1})_{\ell m,\ell' m'}|^2+ 2 Re[\hat{a}^{*}_{\ell m}  ({\Gamma}_R^{-1})_{\ell m,\ell' m'}\hat{a}_{\ell' m'}^{}]}{(1+2\ell)(1+2\ell')}.
    \end{equation}
\end{widetext}

Note that the above expression requires an inversion of the Fisher matrix $\Gamma$, which is often done through the use of SVD regularization. However, the bias introduced by this regularization propagates to the elements of $\hat{A}$ and $K^A$ \cite{Yang_2023}. In order to avoid the effects of this bias on our estimators of angular power spectrum model parameters, we may choose to perform the parameter estimation in the dirty map space, where Fisher matrix inversion is no longer required. 

\subsection{SGWB-EM Angular Cross-Power Spectrum}\label{sec:GWEM-clean}

Cross-correlations of the SGWB anisotropy with anisotropy measured in different EM tracers of structure of matter have the potential to aid in indirectly detecting the anisotropic SGWB \cite{Cusin_2017, Cusin_2019, Yang_2020, Yang_2023}. Such EM tracers include galaxy counts surveys (GC), gravitational weak lensing, and the Cosmic Microwave Background (CMB) \cite{Cusin_2017,Yang_2023}. Each of the EM tracers are expected to be related to different astrophysical and cosmological processes that could give rise to the SGWB. The SGWB-EM cross-correlations, therefore, provide probes of what could give rise to anisotropy in SGWB, as well as of how the matter structure formed.

To estimate the SGWB-EM cross-correlation angular power spectrum in the clean space, we again start with equations Eq. \ref{eq:dirty}, \ref{eq:fisher}, and \ref{eq:p0} to estimate the SGWB spherical harmonics decomposition ($\hat{a}_{\ell m}$'s). On the EM tracer side, the data often come from different surveys binned in a pixel basis on the sky. We then transform the EM data to the spherical harmonic basis and estimate EM tracer spherical harmonics coefficients, ($\hat{b}_{\ell m}$'s), following the process outlined in \cite{HealPix}. By default, $\hat{b}_{\ell m}$'s are already given in the clean space. The EM tracer power spectrum in the clean space, $\hat{B}_\ell$, is given by
\begin{equation}
	\hat{B}_\ell = \frac{1}{(2\ell +1)r_{sky}^2} \sum_{m=-\ell}^\ell \lvert \hat{b}_{\ell m}\rvert^2,
\end{equation}
where $r_{sky}$ is the fraction of the sky that was observed in the survey used for the EM tracer data. Together, the $\hat{a}_{\ell m}$'s and $\hat{b}_{\ell m}$'s combine to construct the SGWB-EM tracer angular cross-correlation power spectrum from the data in the clean space, $\hat{C}_\ell$, given by
\begin{equation}\label{eqn:ClClean_almblm}
	\hat{C}_\ell = \left( \frac{2\pi^2f^3_{ref}}{3H^2_0} \right) \frac{1}{(2\ell +1)r_{sky}} \sum_{m=-\ell}^\ell \left( \hat{a}_{\ell m} \right)^*\hat{b}_{\ell m}.
\end{equation}

As already noted, the $\hat{a}_{\ell m}$'s are multivariate Gaussian distributed with zero mean and covariance given by the regularized inverse Fisher matrix. Currently, the SGWB anisotropy signal-to-noise ratio is very low compared to that obtained in EM surveys. Consequently, the $\hat{b}_{\ell m}$'s contribute negligible noise to the $\hat{C}_\ell$'s. In this limit, we can view Eq. \ref{eqn:ClClean_almblm} as a linear transformation of the multi-variate Gaussian random variables $\hat{a}_{\ell m}$, resulting in multi-variate Gaussian-distributed estimators $\hat{C}_\ell$'s whose covariance matrix $K_C$ can be derived from the inverse Fisher matrix and the $\hat{b}_{\ell m}$'s:
\begin{widetext}
    \begin{equation}\label{eqn:KFisherClean}
    	(K_C)_{\ell,p} = \left( \frac{2\pi^2f^3_{ref}}{3H^2_0} \right)^2 \frac{1}{r^2_{sky}} \frac{1}{(2\ell +1)(2p+1)} \sum_{m,q} \left( \hat{b}_{\ell m}\right)^* \left( \Gamma_R^{-1} \Gamma \Gamma_R^{-1} \right)_{\ell m, pq} \hat{b}_{pq}.
    \end{equation}
\end{widetext}
%
%
We can then proceed to infer model parameters using Gaussian likelihood given by:
\begin{widetext}
    \begin{equation}\label{eqn:lnLike_clean}
        \ln{\mathcal{L}}\left( \hat{C}_\ell \mid C_\ell^{M}(\boldsymbol{\theta}) \right) \propto -\frac{1}{2} \left( \hat{C}_\ell - C_\ell^{M}(\boldsymbol{\theta}) \right) (K_C)^{-1}_{\ell,\ell'} \left( \hat{C}_{\ell'} - C_{\ell'}^{M}(\boldsymbol{\theta}) \right)
    \end{equation}
\end{widetext}
where $C_\ell^{M}(\boldsymbol{\theta})$ is a model cross-power spectrum in the clean space defined by some parameters $\boldsymbol{\theta}$.

\section{Inference in the Dirty Space}\label{sec:Bl}

\subsection{SGWB Angular Power Spectrum in the Dirty Space}\label{sec:dirty_auto}

Theoretical models of the SGWB auto-power angular power spectrum make predictions in the clean space. That is, they are not convolved with any detector response function. Therefore, if we wish to perform a parameter estimation analysis for the angular power spectrum in the dirty map space using some theoretical model, we must convert that model to the dirty space. As shown in Eq. \ref{eq:p0}, the Fisher matrix allows for transformations between the dirty and clean spaces. However, this transformation is done in terms of the individual map elements rather than the angular power spectrum itself. 

As an example, let us assume a simple theoretical model for the clean-space angular power spectrum $A_\ell^M$ as a function of $\ell$:
\begin{equation}\label{eqn:Al_mc_simple}
	A_\ell^{M}(\theta) = \theta \ell.
\end{equation}
Here, $\theta$ is some proportionality constant which we wish to estimate, and the $M$ superscript denotes that the power spectrum is a model. 
%
%
As we show in Appendix \ref{ap:model_auto}, the corresponding model in the dirty map space can be written as
\begin{align}\label{eq:XlM2main}
	 {X}^M_\ell(\theta) =\frac{1}{2\ell+1}\sum_{m=-\ell}^\ell\left(  \sum_{\ell'=0}^{\ell_{max}}A^M_{\ell'}(\theta) \left(\sum_{m'=-\ell'}^{\ell'}|\Gamma_{\ell m, \ell' m'}|^2 \right) \right).
	\end{align}

To demonstrate inference in the dirty space, we perform simulations of both the signal and the noise in the dirty space and apply likelihood maximization. Specifically:

\begin{enumerate}
\item We generate a set of noise-only dirty space elements $\hat{x}_{\ell m}$ by drawing them from a multivariate Gaussian distribution with zero means and the Fisher matrix as the covariance matrix, as done in \cite{o4_aniso}.
\item We next generate clean space model elements $a_{\ell m}^M(\theta_0)$ for a chosen value of the parameter $\theta_0$--- we treat them as independent, Gaussian-distributed random variables. They are complex-valued, and for a given $\ell$ we assume their real and imaginary components may be independently drawn from the Gaussian distribution $\mathcal{N} (0,\frac{1}{2}A^M_\ell(\theta_0))$. For the case when $m = 0$, the corresponding $a_{\ell 0}^M$ are purely real, and they may be drawn from $\mathcal{N} (0,A^M_\ell(\theta_0))$. 
\item We then convert the drawn clean model map into the dirty space by multiplying it with the Fisher matrix:
\begin{equation}\label{eqn:dirtying_alm_mc}
	x_{\ell m}^{M}(\theta_0) = \Gamma_{\ell m, pq} a_{pq}^{M}(\theta_0),
\end{equation}
and we add this dirty signal sky-map to the noise-only map to produce a realistic simulated sky-map:
\begin{align}\label{eqn:signal-inj-dirtymap-elements}
\hat{x}^{sim}_{\ell m}(\theta_0) = \hat{x}_{\ell m} + x^M_{\ell m}(\theta_0). 
\end{align}
The corresponding simulated angular power spectrum is (see Appendix \ref{ap:inject_auto}):
\begin{eqnarray}\label{eq:X_inj_ell}
		\hat{X}_{\ell}^{sim}(\theta_0) & = & \frac{1}{1+2\ell}\sum_{m=-\ell}^{\ell} \left[|\hat{x}^{sim}_{\ell m}(\theta_0)|^2 - \Gamma_{\ell m,\ell m}\right]. \nonumber \\
        &&
\end{eqnarray}
\item We next define the likelihood function to perform statistical inference. While $\hat{X}_{\ell}^{sim}$ follow the generalized multivariate $\chi^2$ distribution, we will approximate it here with a Gaussian, analogously to Eq. \ref{eqn:lnLike_clean_auto}:
\begin{widetext}
    \begin{equation}\label{eqn:lnLike_dirty_auto}
        \ln{\mathcal{L}}\left( \hat{X}_\ell^{sim}(\theta_0) \mid X_\ell^{M}(\theta) \right) \propto -\frac{1}{2} \left( \hat{X}_\ell^{sim}(\theta_0) - X_\ell^{M}(\theta) \right) (K^X(\theta))^{-1}_{\ell,\ell} \left( \hat{X}_\ell^{sim}(\theta_0) - X_\ell^{M}(\theta) \right).
    \end{equation}
\end{widetext}
\item 

For each value of the parameter $\theta$, the dirty-space model spectrum is evaluated using Eq. \ref{eq:XlM2main}. The covariance matrix $(K^X)$ featured in Eq. \ref{eqn:lnLike_dirty_auto} has two contributions. First, there is a contribution due to the detector noise encoded in the Fisher matrix. This contribution is analogous to Eq. \ref{eqn:KFisherClean_auto} and is given by
\begin{widetext}
\begin{equation}\label{eqn:KFisherDirty_auto}
    K^X_{Fisher,\ell, \ell '}=\sum_{m ,m'} \frac{  |( {\Gamma})_{\ell m,\ell' m'}|^2+ 2 Re[\hat{x}^{*}_{\ell m}  {\Gamma}_{\ell m,\ell' m'}\hat{x}_{\ell' m'}^{}]}{(1+2\ell)(1+2\ell')}.
\end{equation}
\end{widetext}
Second, there is a contribution due to the random draw of the signal map elements $a^M_{\ell m}$ in step 2, which introduces statistical fluctuations in the realization of the simulated signal. This contribution to the covariance may be calculated numerically, by repeating steps 2-3 $10^4$ times and drawing $10^4$ realizations of $X_\ell^M$'s for the given value of $\theta$ (see Appendix \ref{ap:model_auto}). One can then  manually calculate the covariance matrix from these samples, which we refer to as $K^X_{Draw}(\theta)$. The total covariance $K^X$ is given as the sum:
\begin{equation}\label{eqn:KFisherDirty_auto}
    	K^X= K^X_{Fisher}+K^X_{Draw}(\theta). 
\end{equation}
Note that, as shown in Appendix \ref{ap:model_auto}, $K^X_{Draw}(\theta)$ depends on the model parameters and therefore must be calculated separately for each set of parameter values when evaluating the likelihood function. 
\end{enumerate}

\subsection{SGWB-EM Angular Cross-Power Spectrum in the Dirty Space}

To develop the inference formalism using the SGWB-EM tracer cross-correlation angular power spectrum in dirty space, we follow a similar procedure to the SGWB auto-power case. On the model side, 
we again use a simple theoretical model for the clean-space SGWB-EM angular power spectrum $C_\ell^M$ as a function of $\ell$:
\begin{equation}\label{eqn:Cl_mc_simple}
    C_\ell^M (\rho) = \rho \ell \sqrt{A_\ell^M B_\ell^M}.
\end{equation}
Here, $A_\ell^M$'s represent the SGWB clean-space angular power spectrum, which we assume is known and can be calculated using Eq. \ref{eqn:Al_mc_simple} with $\theta=5\times 10^{-99}$. We also assume that the EM tracer angular power spectrum is known and given by $B_\ell^M = A_\ell^M/1000$. Finally, $\rho$ is the correlation coefficient between SGWB and the EM tracer, and we treat it as a free parameter in our inference framework.  

As we show in Appendix \ref{app:deriv-of-analytical-model-Zl}, the corresponding dirty-space model can be written analytically as:
\begin{equation}
    Z_\ell^M (\rho) = \frac{1}{2\ell+1}  \sum_{m=-\ell}^\ell  \sum_{\ell'=0}^{\ell_{max}}  \sum_{m'=-\ell'}^{\ell'} \Gamma_{\ell m,\ell ' m'}  C_{\ell'}^M (\rho) \Gamma_{\ell'm', \ell m}
\end{equation}

Similarly to the auto-power case, we demonstrate the inference formalism using simulated data. 
We follow a similar procedure to that outlined in Section \ref{sec:dirty_auto}, with slight complications due to the inclusion of the EM tracer. We assume that the EM tracer clean map elements, $b_{\ell m}^M$, are complex-Gaussian distributed random variables with mean zero. Similarly to the auto-power case, for a given $\ell$ we assume that the real and imaginary components of $a_{\ell m}^M$ and $b_{\ell m}^M$ can be drawn from a bi-variate Gaussian distribution $\mathcal{N}(\boldsymbol{0},\Sigma_{ab})$ where $\Sigma_{ab}$ is given by
\begin{equation}\label{eqn:Sigma_ab}
    \Sigma_{ab} = \begin{pmatrix}
        \frac{1}{2}A_\ell^M & \frac{1}{2}C_\ell^M (\rho_0) \\
        \frac{1}{2}C_\ell^M (\rho_0) & \frac{1}{2}B_\ell^M
    \end{pmatrix},
\end{equation}
except for the $m=0$ case which is purely real:
\begin{equation}
    \Sigma_{ab}^{m=0} = \begin{pmatrix}
        A_\ell^M & C_\ell^M (\rho_0) \\
        C_\ell^M (\rho_0) & B_\ell^M
    \end{pmatrix}.
\end{equation}
Here, $\rho_0$ is the correlation coefficient chosen for the signal simulation.
The SGWB and EM tracer sky-maps drawn from this distribution are transformed into the dirty space, using Eq. \ref{eqn:dirtying_alm_mc} for the SGWB map elements and the following equation for the EM tracer map elements:
\begin{equation}
     y_{\ell m}^M (\rho_0) = \sum_{\ell'=0}^{\ell'=\ell_{max}} \sum_{m'=-\ell'}^{m'=\ell'} \Gamma_{\ell m,\ell'm'}b_{\ell'm'}^M (\rho_0)
\end{equation}

The dirty-space simulated SGWB sky-map is then given by Eq. \ref{eqn:signal-inj-dirtymap-elements}, and we define the simulated SGWB-EM tracer angular cross-power spectrum in dirty space as:
\begin{equation}
    \hat{Z}^{sim}_\ell (\rho_0) = \frac{1}{2\ell+1} \sum_{m=-\ell}^\ell \left( \hat{x}^{sim}_{\ell m} (\rho_0) \right)^* y_{\ell m}^M (\rho_0)
\end{equation}

For parameter estimation in the dirty space, the log-likelihood is analogous to Eq. \ref{eqn:lnLike_clean}:
\begin{widetext}
    \begin{equation}\label{eqn:lnLike-Zl}
	\ln{\mathcal{L}\left( \hat{Z}_\ell^{sim} (\rho_0) \mid Z_\ell^{M}(\rho) \right)} \propto -\frac{1}{2} \left( \hat{Z}_\ell^{sim} (\rho_0) - Z_\ell^{M}(\rho) \right) \left( K^Z\right) ^{-1}_{\ell, \ell} \left( \hat{Z}_\ell^{sim} (\rho_0) - Z_\ell^{M}(\rho) \right)
 \end{equation}
\end{widetext}

The covariance matrix $K^Z$ has two contributions, just like in the SGWB-only case. First, there is a contribution due to the detectors' noise propagated from the Fisher matrix as:
\begin{widetext}
    \begin{align}\label{eqn:KFisherDirty}
	(K_{Fisher}^Z)_{\ell ,\ell'} = \frac{1}{2\ell+1} \frac{1}{2\ell'+1} \sum_{m=-\ell}^\ell \sum_{m'=-\ell'}^{\ell'} \left( \hat{y}_{\ell m} \right)^* \hat{y}_{\ell'm'} \frac{1}{2} &\left[ \Gamma_{\ell m, \ell'm'} - (-1)^m \Gamma_{\ell -m, \ell'm'}\right. \\
   &\left.+ (-1)^{m'} \Gamma_{\ell m, \ell'-m'} + (-1)^{m+m'} \Gamma_{\ell -m, \ell'-m'} \right]. \nonumber
\end{align}
\end{widetext}

Second, there is an additional uncertainty due to the random draw of the signal map elements $a_{\ell m}^M$ and $b_{\ell m}^M$. This contribution is again calculated numerically by drawing SGWB and EM tracer map elements and calculating $Z_\ell^M$ 1,000 times (see Appendix \ref{app:crossspower-Kdraw} for more information on this calculation). We refer to this covariance as $K^Z_{draw}$, to give the total covariance as sum of the two sources of covariance:
\begin{equation}
    K^Z (\rho) = K^Z_{Fisher} +K^Z_{draw}(\rho).
\end{equation}

\section{Results}\label{sec:Results}

Here we apply the methodology described in the preceding sections and report the results of our parameter estimation scheme in the dirty map space for both the SGWB auto-power case as well as the cross-power case. 

\subsection{SGWB Auto-Power Results}
We carry out the procedure outlined in Sec. \ref{sec:dirty_auto}. We add simulated model signal into noise-only maps and obtain dirty space angular power spectra which we then compare to the model using the likelihood defined in Eq. \ref{eqn:lnLike_dirty_auto}. Noise maps are generated using the Fisher matrix $\Gamma$ from Advanced LIGO's third observing run (O3). The likelihood function is evaluated over a grid of values of the parameter $\theta$ centered on the true injected value $\theta_0$. In a Bayesian formalism, this choice constitutes a uniform prior for our model parameter, and the posterior distribution of $\theta$ is, therefore, equal to the likelihood function up to a normalization factor. 

\begin{figure}
    \centering
    \includegraphics[width=0.98\linewidth]{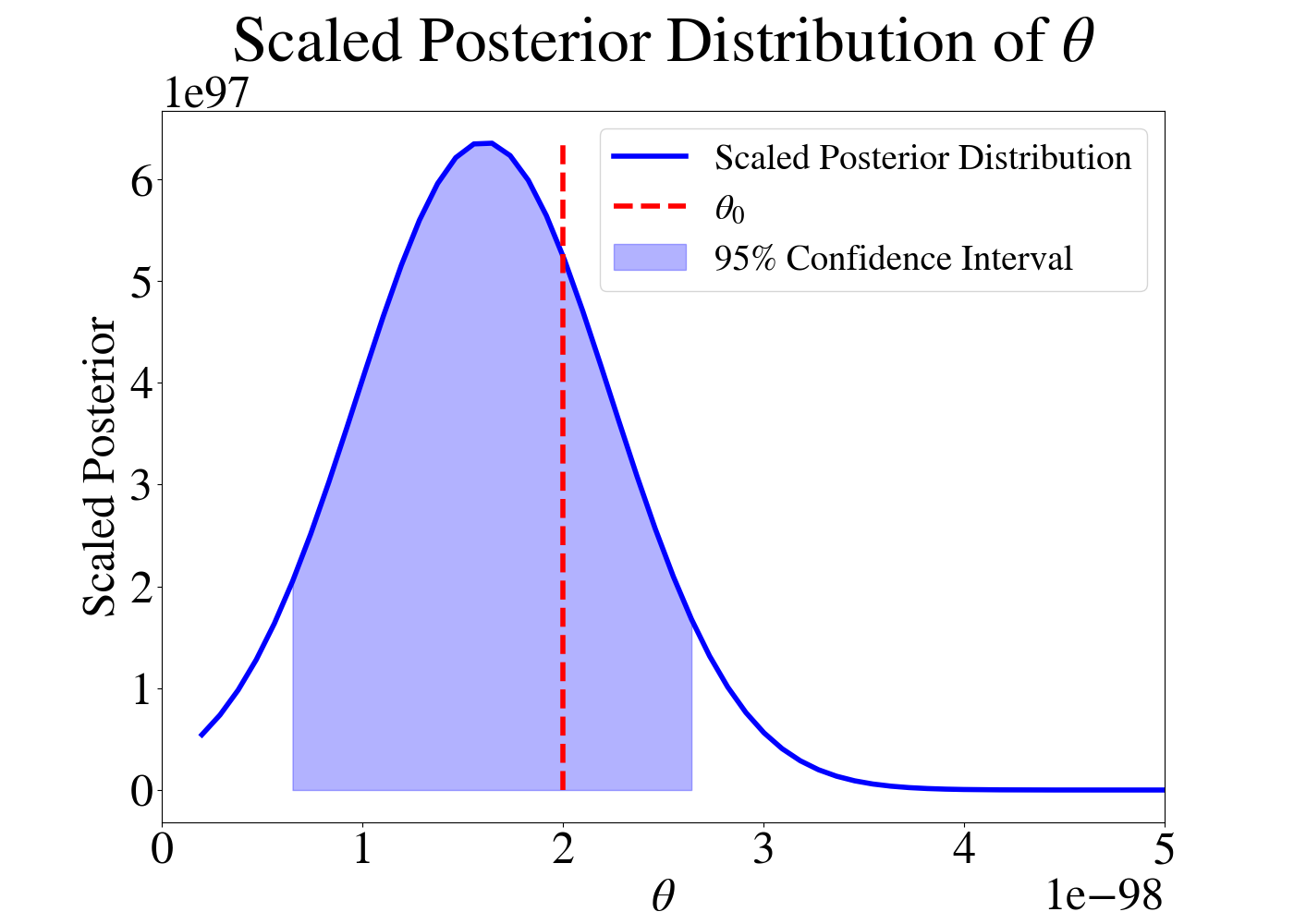}
    \caption{Example of a posterior distribution of the model parameter $\theta$ with $\ell_{max}=6$ and simulated value $\theta_0 = 2 \times 10^{-98}$ with a 95\% confidence interval.}
    \label{fig:auto-posterior}
\end{figure}

\begin{figure}
    \centering
    \includegraphics[width=0.98\linewidth]{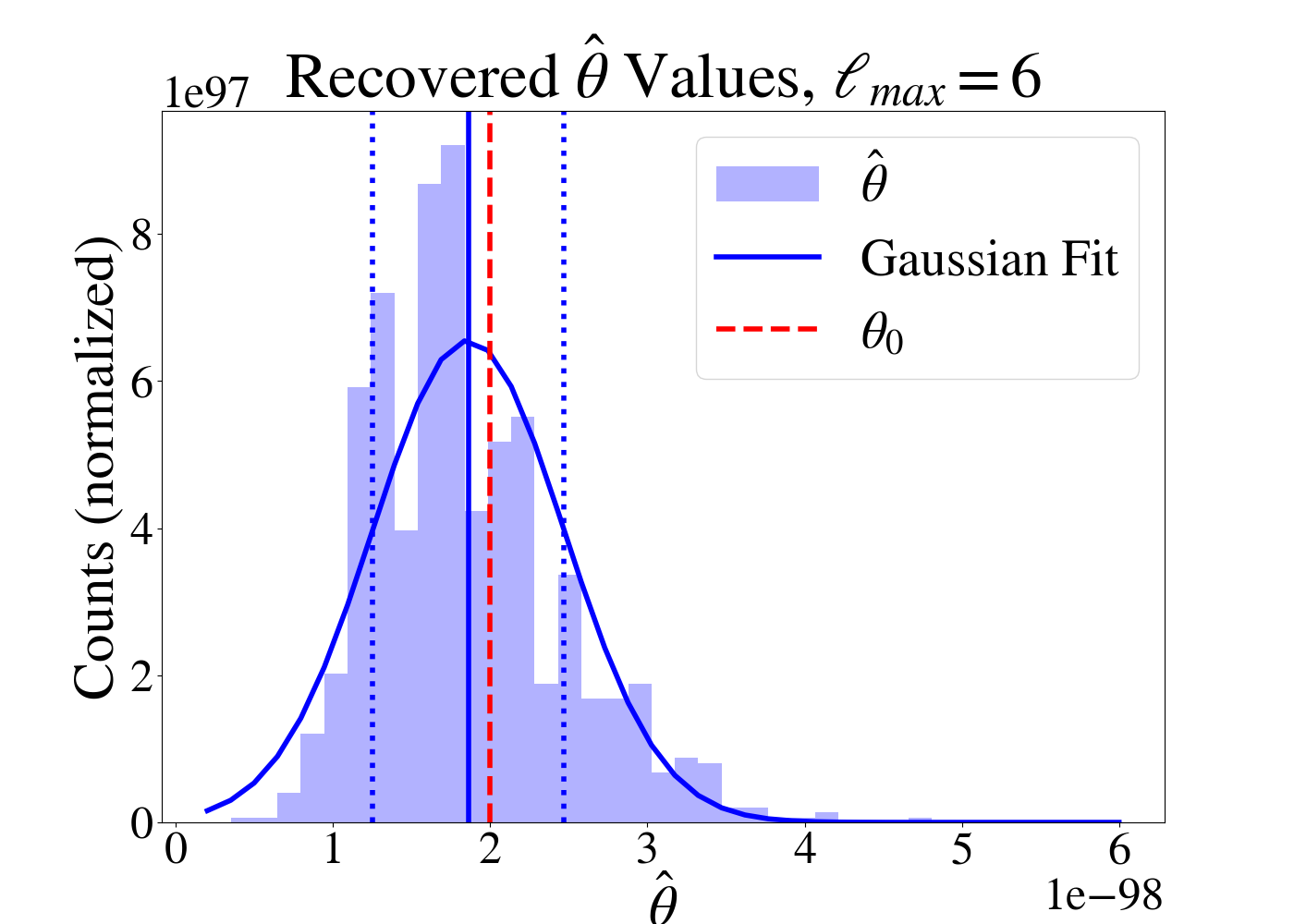}
    \caption{Histogram of recovered $\theta$ values at $\theta_0 = 2 \times 10^{-98}$ and $\ell_{max}=6$. A Gaussian fit is used to determine the estimator $\hat{\theta}$ (see text).}
    \label{fig:auto-hist-recov-p}
\end{figure}

In Fig. \ref{fig:auto-posterior} we show an example of a posterior distribution for the case where $\theta_0 = 2 \times 10^{-98}$ and $\ell_{max}=6$. The value of $\theta$ for which the posterior is at its peak is recorded, and the process is repeated 1000 times, each with a different realization of the model signal added into a new realization of noise. The collection of 1000 recovered $\hat\theta$ values are binned into a histogram shown in Fig. \ref{fig:auto-hist-recov-p}. We choose to approximate this distribution of recovered $\hat\theta$ values as Gaussian and apply a Gaussian fit. The mean of this fit $\mu_\theta$ and its standard deviation $\sigma_\theta$ then capture how well our inference formalism recovers the true model parameter $\theta_0$. 

\begin{figure}
    \centering
    \includegraphics[width=0.98\linewidth]{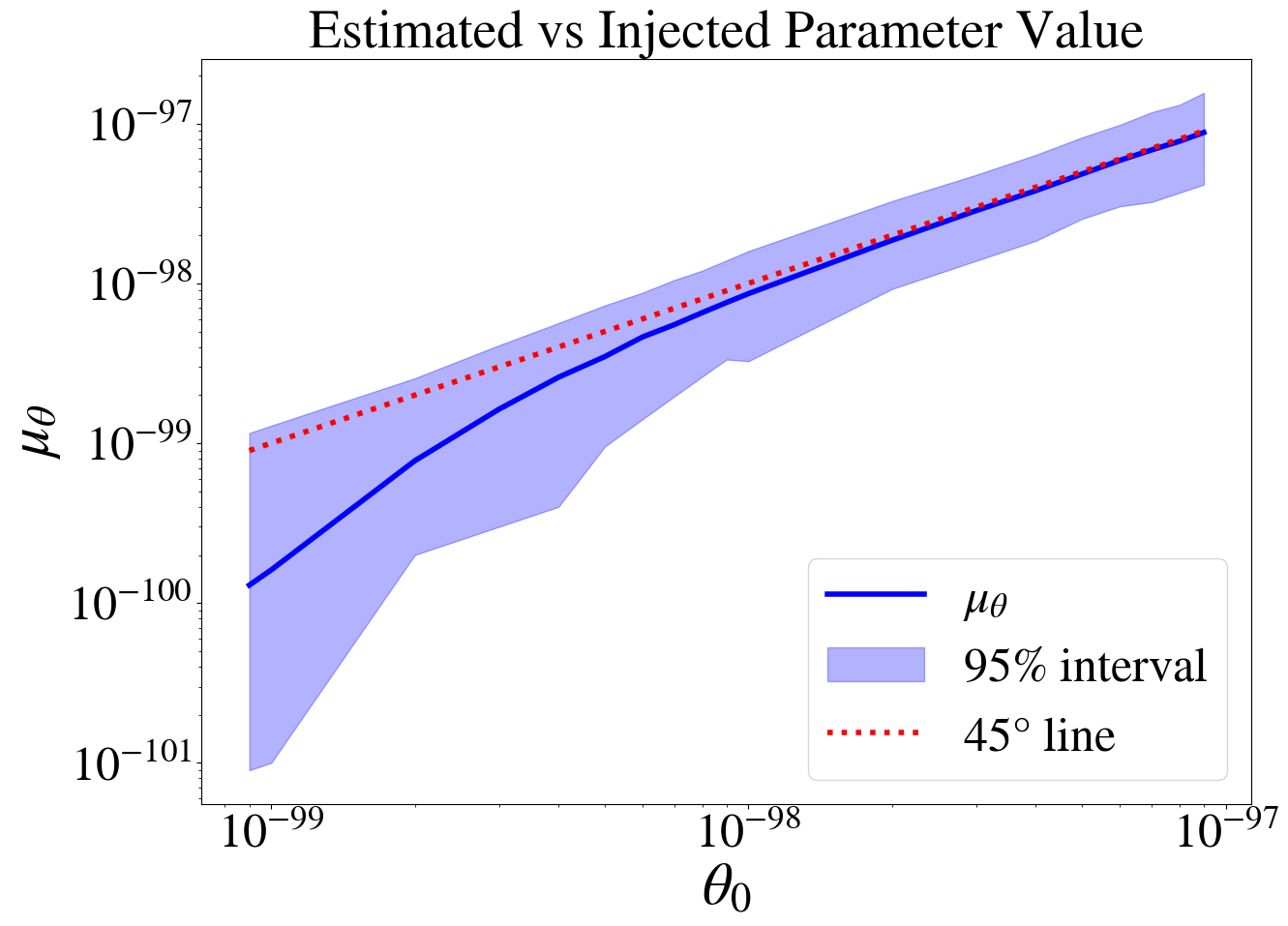}
    \caption{The 95\% credible interval of the recovered $\hat\theta$ values is shown for increasing $\theta_0$. The red line indicates a perfect recovery as the 45$\degree$ line.} 
    \label{fig:auto-varying-injected-signal}
\end{figure}

We plot $\mu_\theta$ as a function of the injected parameter $\theta_0$ in Fig. \ref{fig:auto-varying-injected-signal}. For a weak signal injection (low $\theta_0$), the simulated data are dominated by the detector noise, and our mean recovery deviates from the correct parameter value, obtaining estimators which fall below $\theta_0$. This bias at low $\theta_0$ likely comes from the assumption of a Gaussian fit in Fig. \ref{fig:auto-hist-recov-p}. As the strength of the injected signal increases, we obtain estimators which fall increasingly close to the true value. 

Note however that the 95$\%$ credible interval shown in Fig. \ref{fig:auto-varying-injected-signal} increases in width as the signal gets stronger. This is because the contribution to covariance $K^X_{Draw}$ depends on the model parameters and therefore increases when the signal is stronger (c.f. Eq. \ref{eqn:KFisherDirty_auto}). Injecting an arbitrarily strong signal into detector noise (characterized by $K^X_{Fisher}$) therefore requires introducing a correspondingly large amount of uncertainty from $K^X_{Draw}$. This latter contribution dominates the detector uncertainty for strong signals. While this contribution originates from our procedure of artificially adding signal into noise maps, there is an analogous source of variance which arises in real data analysis (without any signal simulations). This so-called cosmic variance results from only having a single realization of the SGWB and is proportional to the square of the angular power spectrum itself \cite{Allen_2023, Bernardo_2022, 2016ApJ...819..163R, Ravi_2012, Cornish_2013}. 

\begin{figure}[h!]
    \centering
    \includegraphics[width=0.98\linewidth]{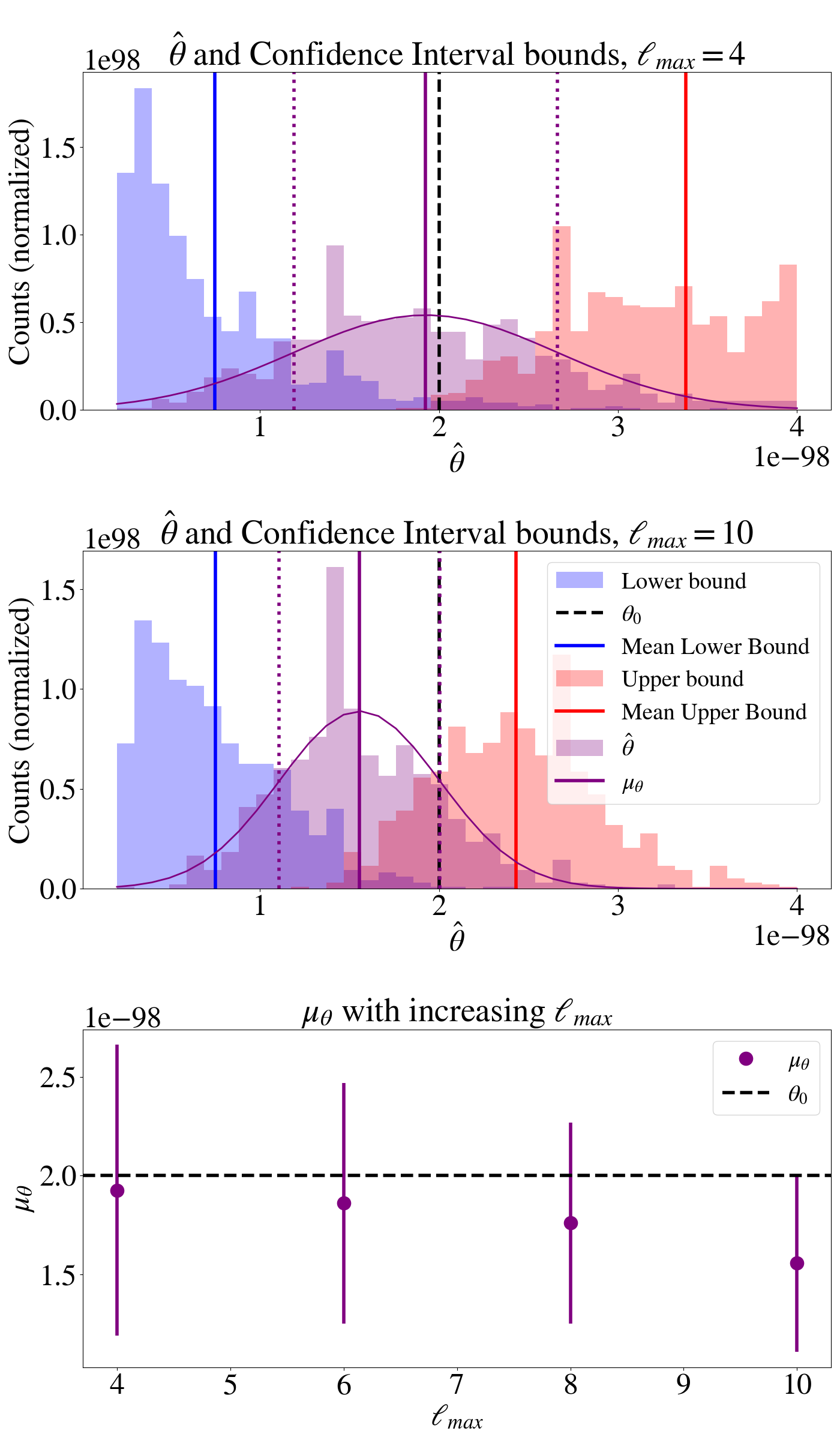}
    \caption{Increasing $\ell_{max}$ decreases the variance of the gaussian fit, all recovering $\theta_0$ (black dashed line) within uncertainty. \textit{Purple:} Histogram of recovered $\theta$ with a Gaussian fit and vertical lines indicating mean and $\sigma$. \textit{Red:} Distribution of the upper bounds of 95\% confidence intervals with the mean indicated. \textit{Blue:} Distribution of the lower bounds of 95\% confidence intervals with the mean indicated. \textit{Top:} histograms and Gaussian fit for $\ell_{max}=4$. \textit{Middle:} histograms and Gaussian fit for $\ell_{max}=10$. \textit{Bottom:} the mean recovered $\mu_\theta$ with $\sigma_\theta$ as the error bars. Each point was generated from 1000 posterior distributions of $\theta$. }
    \label{fig:auto-increasing-lmax}
\end{figure}

We may also see how our ability to estimate our model parameter is affected by our choice of $\ell_{max}$. In Fig. \ref{fig:auto-increasing-lmax} we show histogram plots for $\ell_{max}=4$ and $\ell_{max}=10$. In each plot there are three histograms showing the recovered $\theta$ values along with the upper and lower bounds of their corresponding 95$\%$ credible intervals. For $\ell_{max}=4$ the recovered values peak close to the true value $\theta_0$, but the distributions of recovered values and confidence interval bounds are widely spread, suggesting more variability in the corresponding posterior distributions. When $\ell_{max}=10$ we see tighter distributions of recovered values and confidence interval bounds, but our estimator $\hat{\theta}$ is slightly biased towards values below $\theta_0$. 
This trend is illustrated in the bottom plot of Fig. \ref{fig:auto-increasing-lmax}, where increasing $\ell_{max}$ decreases uncertainty in the recovered $\theta$ but leads to slight under-estimates relative to $\theta_0$. 
At high $\ell_{max}$ each $X^{sim}_\ell$ is an average over a larger number of map elements $x^{sim}_{\ell m}$ (c.f. Eq. \ref{eq:X_inj_ell}), leading to more tightly distributed $X^{sim}_\ell$ across different realizations of noise and a correspondingly smaller uncertainty $\sigma_\theta$. However, higher-order multipoles will include directions in the sky to which the Advanced LIGO detectors are less sensitive, increasing the variance contribution from $K^X_{Fisher}$.

\subsection{SGWB-EM Tracer Cross-Power Results}

\begin{figure}
    \centering
    \includegraphics[width=0.98\linewidth]{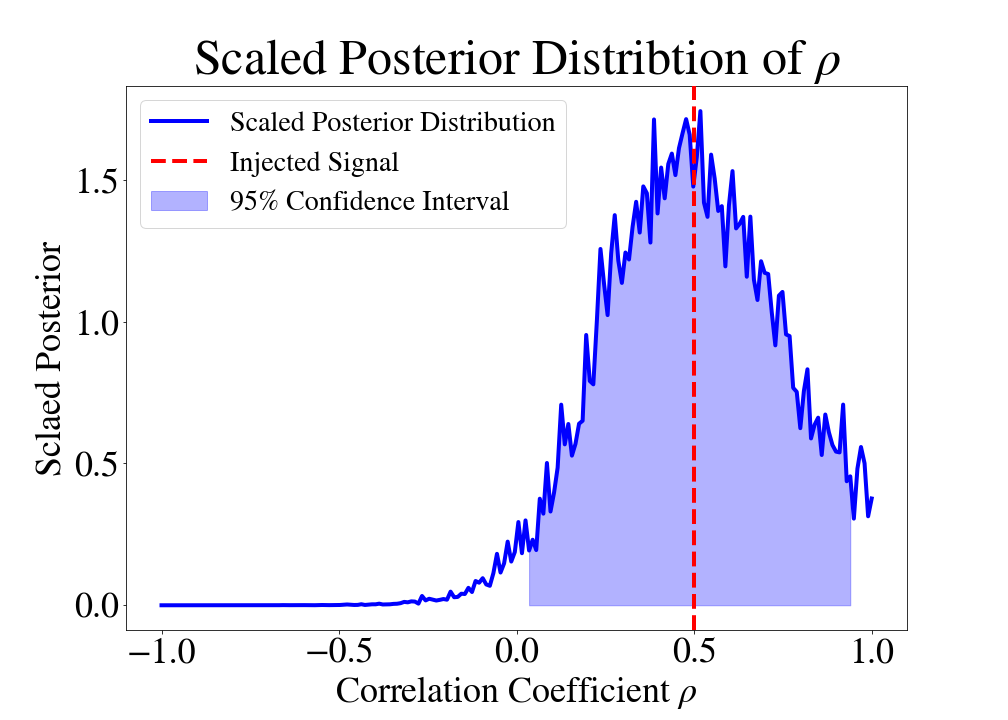}
    \caption{Posterior distribution of the recovered correlation coefficient $\rho$ is shown along with the recovered peak value $\hat\rho = 0.5$, for $\ell_{max}=6$. The 95\% confidence interval of 0.1-0.9 includes the true $\rho_0=0.5$ value. }
    \label{fig:gwem-posterior}
\end{figure}

\begin{figure}
    \centering
    \includegraphics[width=0.98\linewidth]{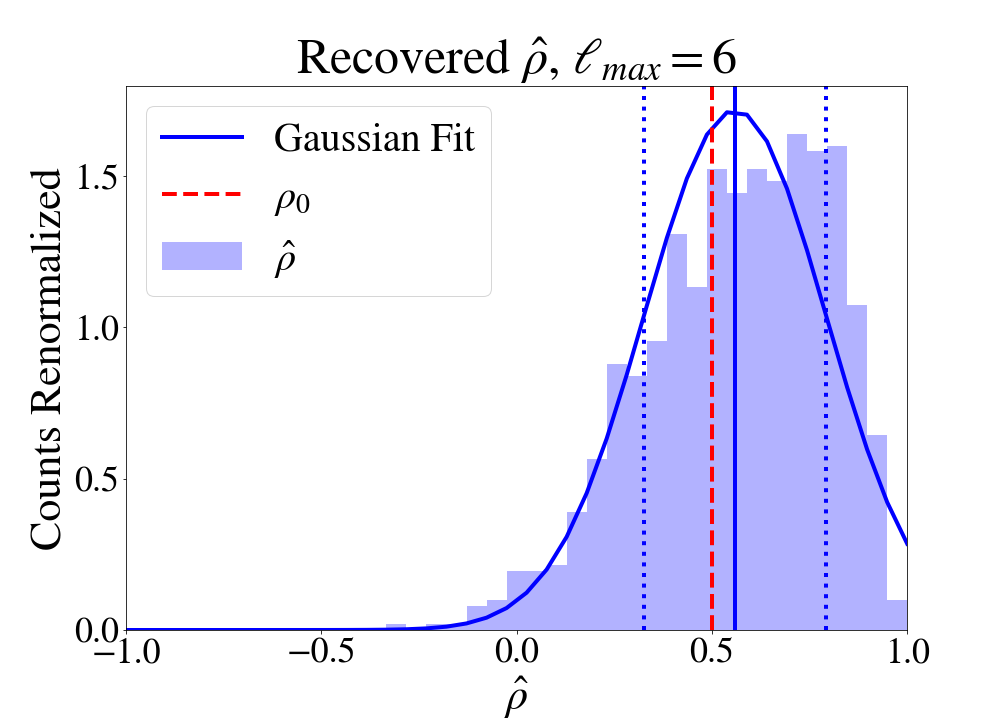}
    \caption{Histogram of the recovered $\hat\rho$ (blue) with a Gaussian fit indicating a mean $\mu_\rho = 0.6$ and a $\sigma_\rho=0.2$ which aligns with a $\rho_0=0.5$ (red), $\ell_{max}=6$.}
    \label{fig:gwem-hist-recov-p}
\end{figure}

\begin{figure}
    \centering
    \includegraphics[width=0.98\linewidth]{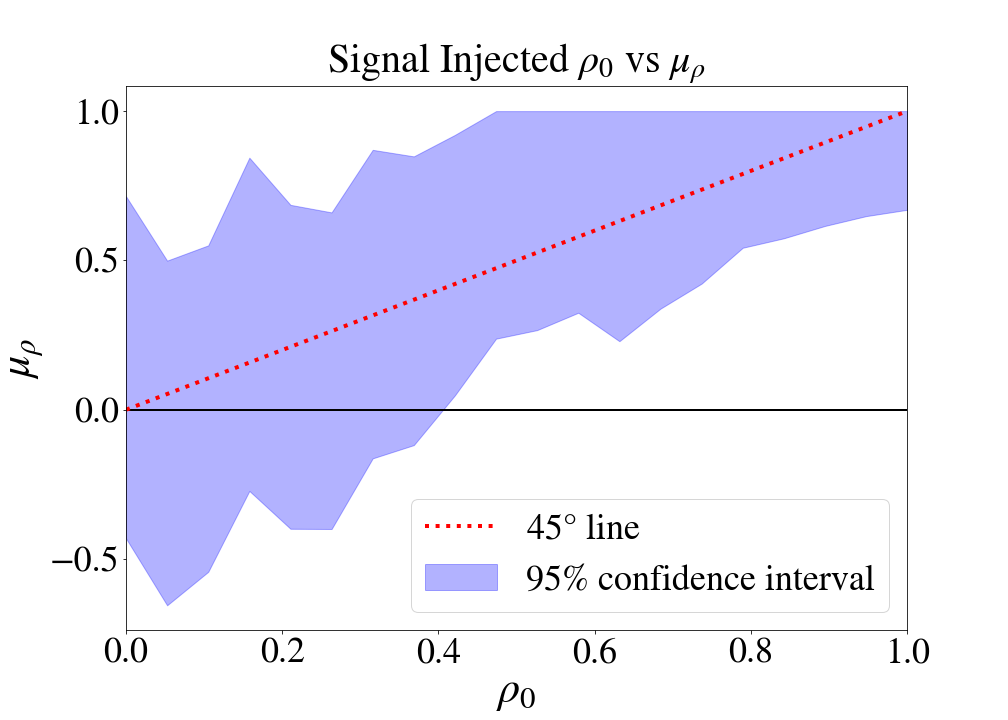}
    \caption{The 95\% confidence interval of the recovered $\hat\rho$ vs $\rho_0$ is shown, with the red line indicating a perfect recovery as the $45\degree$ line, $\ell_{max}=6$. A truncated normal was fit to the distributions and used to estimate the 95\% confidence interval.}
    \label{fig:gwem-varying-injected-signal}
\end{figure}

\begin{figure}
    \centering
    \includegraphics[width=0.98\linewidth]{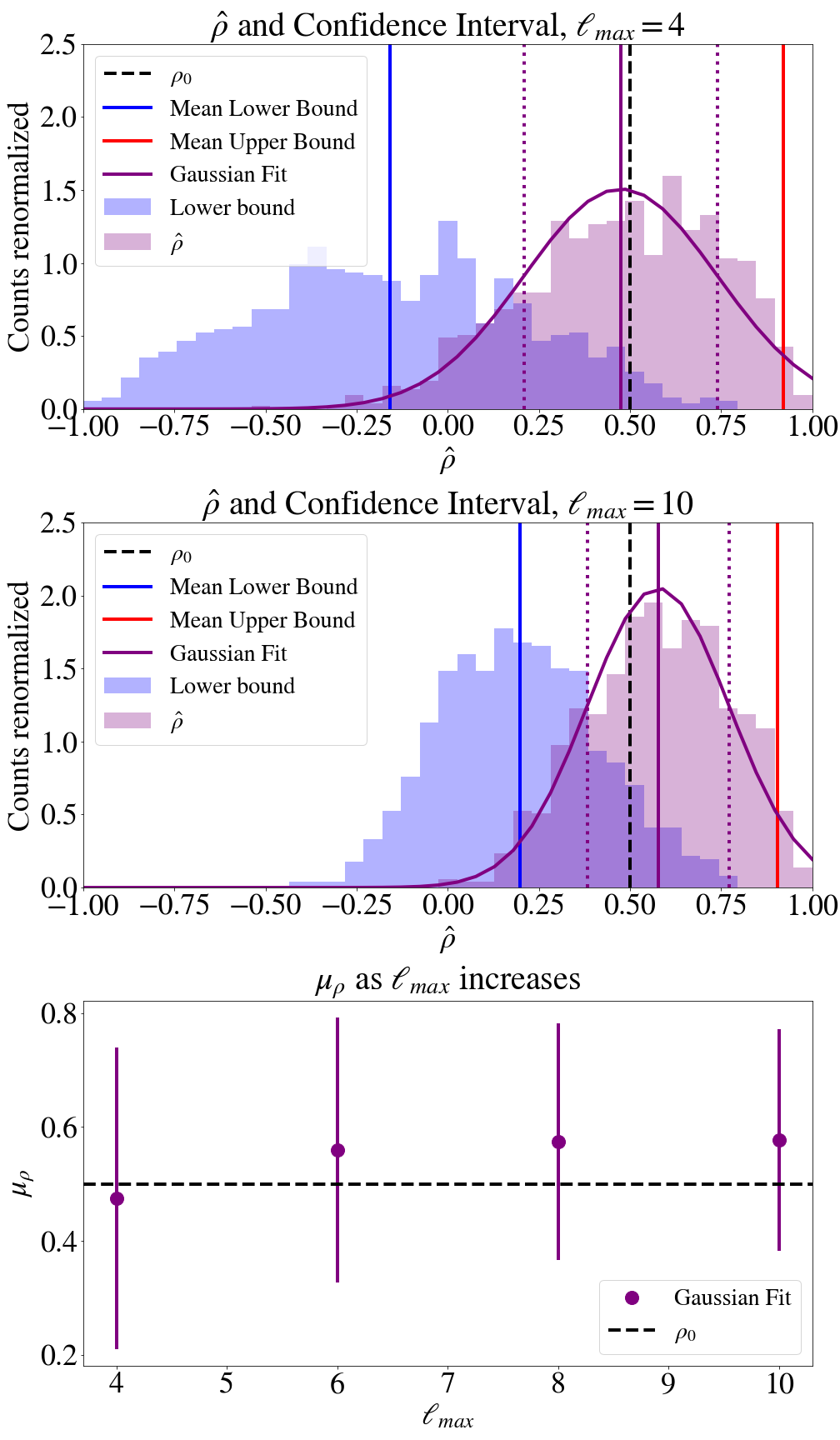}
    \caption{Increasing $\ell_{max}$ decreases the variance of the gaussian fit, all recovering $\rho_0$ (black dashed line) within uncertainty. \textit{Purple:} Histogram of recovered $\rho$ with a Gaussian fit and vertical lines indicating mean and $\sigma$. \textit{Red:} Mean of the upper bounds of 95\% confidence intervals. \textit{Blue:} Distribution of the lower bounds of 95\% confidence intervals with the mean indicated. \textit{Top:} histograms and Gaussian fit for $\ell_{max}=4$. \textit{Middle:} histograms and Gaussian fit for $\ell_{max}=10$. \textit{Bottom:} the mean recovered $\mu_\rho$ with $\sigma_\rho$ as the error bars. Each point was generated from 1000 posterior distributions of $\rho$.
    }
    \label{fig:gwem-increasing-lmax}
\end{figure}

Using the simple linear model for the SGWB, EM tracer, and SGWB-EM tracer angular power spectra described by Eqs. \ref{eqn:Al_mc_simple} and \ref{eqn:Cl_mc_simple}, we simulated a signal with a correlation coefficient of $\rho_0=0.5$ and added it to the LIGO O3 noise with an $\ell_{max}=6$. 
Using a step size in $\rho$ of 0.01 from -1 to 1 and using $N_{draw}=1000$ draws to compute $K_{draw}^Z$, we performed a grid search to produce the posterior displayed in Fig. \ref{fig:gwem-posterior}. The mean of the posterior distribution is $\hat\rho= 0.5$, recovering well the simulated value $\rho_0$, with a 95\% confidence interval of 0.1-0.9. The width of the posterior distribution allows for a determination of the correlation coefficient with large error bars, mainly indicating whether or not the power spectra are correlated. 

This is also demonstrated in Fig. \ref{fig:gwem-hist-recov-p}, where the posterior distribution was generated 1000 times, each time using a different realization of noise and the $\rho_0=0.5$ injected signal, and the mean value of the posterior distribution $\hat\rho$ was added to the histogram displayed. 
The mean, $\mu_\rho$, and standard deviation, $\sigma_\rho$, of the histogram in Fig. \ref{fig:gwem-hist-recov-p} when fitted with a Gaussian distribution, are 0.6 and 0.2 respectively. 
Even though the estimate of $\rho$ suffers from large uncertainty, all the recovered means are positively correlated, allowing us to draw the conclusion that there is a correlation recovered between the SGWB and EM tracer angular power spectra. 

Repeating the process for Fig. \ref{fig:gwem-hist-recov-p} while varying the $\rho_0$ from 0 to 1, Fig. \ref{fig:gwem-varying-injected-signal} displays the recovered $\mu_\rho$ and 95\% confidence interval at each $\rho_0$. As $\rho_0$ increases, the uncertainty of the recovered $\rho$ decreases. For $\rho_0 > 0.4$, we are able to detect the signal at 95\% confidence. Note that as the injected signal increases, the posterior distribution is truncated by the limitations of the parameter space. Hence, we fit a truncated normal distribution to extract the 95\% confidence interval displayed in Fig. \ref{fig:gwem-varying-injected-signal}. 

As in the SGWB auto-power case, increasing $\ell_{max}$ reduces the width of the $\rho$ posterior distribution. The top two panels of Fig. \ref{fig:gwem-increasing-lmax} directly compare the posterior distributions for $\ell_{max} = 4$ and 10. As in the SGWB auto-power case, $\ell_{max} = 10$ yields significantly tighter posterior, and both are well consistent with the true value of the simulated parameter $\rho_0 = 0.5$. The third panel of Fig. \ref{fig:gwem-increasing-lmax} shows the behavior of $\mu_\rho$ (with the corresponding $\sigma_\rho$ denoted as error bars) as a function of $\ell_{max}$. As  $\ell_{max}$ increases from $\ell_{max} = 4$ to $\ell_{max} = 10$, $\sigma_\rho$ decreases by 33\%, roughly consistent with the naive expected improvement due to the increased number of data points, $1-\sqrt{\frac{4}{10}}=0.37$. 

As $\ell_{max}$ increases, the upper bound of the confidence interval remains relatively unchanged due to the parameter space cutoff at $\rho=1$. However, the lower bound, which initially shows a wide Gaussian distribution, narrows. This suggests that at higher $\ell_{max}$ values, a single trial posterior is more representative of the true distribution. The increase in spherical harmonic coefficients drawn for $C_\ell$ at higher $\ell$ values ($2\ell+1$ spherical harmonic coefficients per $\ell$ value), reduces the impact of draw uncertainty on each trial posterior, resulting in the trend seen in Fig. \ref{fig:gwem-increasing-lmax}.


\section{Conclusion}\label{sec:Results}

SGWB auto-correlation searches for the anisotropic SGWB continue to be a promising avenue of research given their potential for informing our understanding of the distribution of GW sources both locally within our galaxy and also in the large-scale structure of the universe.
Cross-correlation of the SGWB with EM tracers of structure in the universe is a powerful method that simultaneously probes the sources or anisotropy in the SGWB and has the possibility of indirect detection of the anisotropic SGWB before direct detection methods.
Parameter estimation for these analyses are hindered by angular resolution limits stemming from inverting an ill-defined Fisher matrix.

We provided an alternative method of parameter estimation, for the auto-power and cross-power cases, that avoids inversion of the ill-defined Fisher matrix.
We achieved this by convolving the power spectra models with the GW detector response such that the analysis is performed in the dirty space.
We also demonstrated the effectiveness of this framework with simple models and signals injected into noise simulated using LIGO O3 sensitivity.

In the GW auto-power case, we have demonstrated our ability to perform parameter estimation for a simple model with a single parameter. As the strength of the injected signal increased, our parameter estimates increased in accuracy, but the width of our confidence intervals increased proportionally to signal strength. We tested our parameter estimation for spherical harmonic multipoles of $\ell_{max}=4$ up to $\ell_{max}=10$ and found that, while the precision of our parameter estimates increased at higher $\ell_{max}$, their accuracy slightly decreased.

For the cross-power case, using the correlation coefficient $\rho$ as our free parameter, we were able to demonstrate that we can detect a correlation between SGWB and EM tracer sky map distributions down to a correlation of $\rho_0=0.4$ for $\ell_{max}=6$ with increasing sensitivity as $\ell_{max}$ is increased.

There are some limitations to this framework for parameter estimation. 
For both cases, we used simple models that we expect would be relatively easy to recover and interpret. 
Increasing complexity by increasing $\ell_{max}$, including a more complicated model, or estimating more parameters proved to be computationally expensive under the current code framework.

Another limitation was the draw uncertainty, as it appeared to be the dominant source of uncertainty. 
We posit this is a good representation of the cosmic variance, but it nevertheless limits the sensitivity of the analysis for both the auto-power and cross-power cases. As we increase the angular scale, this affect decreases.

In this study we have made the simplifying assumption that the SGWB auto-power angular spectrum is approximated as a Gaussian-distributed random variable, allowing us to define a simple likelihood function for our parameter estimation. While the assumption of Gaussianity is justified for high values of $\ell_{max}$ (as is done for other contexts such as the cosmic microwave background \cite{Netterfield_2002, PhysRevLett.85.1366, Piacentini_2006, Jungman_1996, Knox_1995}), it is less appropriate at the larger angular scales to which the SGWB search is limited \cite{Agarwal_2023, Bellomo_2022}. It has been shown that the angular power spectrum is more accurately described by a generalized $\chi^2$ distribution for which a closed-form likelihood does not exist, and distribution-free model validation methods have been developed to circumvent this issue \cite{algeri_2025, zhang_2025}.


Future work in this area could include applying this framework to various proposed theoretical models. Such theoretical models for an anisotropic SGWB due to CBC events may be found in \cite{Cusin:2017mjm, Cusin:2019jpv} for the auto-power case and \cite{Yang_2023} for the cross-power case.

An aspect we did not explore is the impact of shot noise to this framework for parameter estimation as is included for the cross-power case in \cite{Yang_2023}. 
This is the contribution to variance which arises from the spatial and temporal discreteness of signal sources, and future work should include it in the covariance calculations.

For the cross-power case, future work should include the EM tracer sky maps for GC, weak lensing, and the CMB into the analysis. Further, more sophisticated models should be implemented in place of the simple toy model used in this study.


\begin{acknowledgements}
    The authors thank Daniel Warshofsky for his assistance in increasing the code efficiency of the SGWB-EM cross-power simulations. 
    The authors are grateful for computational resources provided by the LIGO Laboratory and supported by National Science Foundation Grants PHY-0757058 and PHY-0823459.
\end{acknowledgements}

\appendix

\section{Dirty space SGWB auto-power angular power spectrum estimators}
\label{ap:estimator_auto}

Consider the estimators of the clean map elements $\hat{a}_{\ell m}$, dirty map elements $\hat{x}_{\ell m}$, and Fisher matrix $\Gamma_{\ell m \ell' m'}$. From \cite{sph_methods} we have, in the weak signal approximation,

\begin{align}\label{eq:Exp1}
    \langle \hat{x}_{\ell m}\hat{x}^*_{\ell'm'}\rangle - \langle \hat{x}_{\ell m}\rangle \langle \hat{x}^*_{\ell'm'}\rangle \approx \Gamma_{\ell m \ell' m'}
    \\
    \langle \hat{a}_{\ell m}\hat{a}^*_{\ell'm'}\rangle - \langle \hat{a}_{\ell m}\rangle \langle \hat{a}^*_{\ell'm'}\rangle \approx \Gamma^{-1}_{\ell m \ell' m'}
    \end{align}

If $\ell = \ell'$ and $m = m'$, then we have
    \begin{align}\label{eq:Exp3}
    \langle |\hat{x}_{\ell m}|^2 \rangle - \langle \hat{x}_{\ell m}\rangle^2  \approx \Gamma_{\ell m \ell m} = var(\hat{x}_{\ell m})
    \end{align}
   \begin{align}\label{eq:Exp4}
    \langle |\hat{a}_{\ell m}|^2 \rangle - \langle \hat{a}_{\ell m}\rangle^2  \approx \Gamma^{-1}_{\ell m \ell m} = var(\hat{a}_{\ell m})
    \end{align}
Now consider the (biased) estimators of the clean map angular power spectrum given by \cite{sph_methods}:

	\begin{align}
	 \hat{A}_\ell = \frac{1}{2\ell+1}\sum_m |\hat{a}_{\ell m}|^2.               
	\end{align}
Find the expectation value of both sides:
	\begin{align}
	 \langle \hat{A}_\ell \rangle= \langle \frac{1}{2\ell+1}\sum_m |\hat{a}_{\ell m}|^2\rangle = \frac{1}{2\ell+1}\sum_m \langle|\hat{a}_{\ell m}|^2\rangle.               
	\end{align}
Substitute from Eq. \ref{eq:Exp4}:
	\begin{align}\label{eq:Al1}
	 \langle \hat{A}_\ell \rangle &=\frac{1}{2\ell+1}\sum_m \left( \langle\hat{a}_{\ell m}\rangle^2+\Gamma^{-1}_{\ell m \ell m} \right) \\ &= \frac{1}{2\ell+1}\sum_m\Gamma^{-1}_{\ell m \ell m} + \frac{1}{2\ell+1}\sum_m\langle\hat{a}_{\ell m}\rangle^2.               
	\end{align}
If the purpose of this estimator is to estimate the angular power spectrum of the SGWB sky realization in our universe, then we can assume $\langle\hat{a}_{\ell m}\rangle \neq 0$. Note that this is different from estimating the {\it true} SGWB angular power spectrum, of which our universe is only one realization---this is further discussed below. Therefore, we can define
	\begin{align}\label{eq:trueAl}
	 {A}_\ell = \frac{1}{2\ell+1}\sum_m \langle\hat{a}_{\ell m}\rangle^2,              
	\end{align}
as the angular power spectrum in our universe, and we can rearrange  Eq. \ref{eq:Al1} to give the result from \cite{sph_methods}:
	\begin{align}\label{eq:Al2}
	 \langle \hat{A}_\ell \rangle = \frac{1}{2\ell+1}\sum_m\Gamma^{-1}_{\ell m \ell m} + A_\ell.               
	\end{align}
	\begin{align}\label{eq:Al2}
	  \hat{A}'_\ell  =  \hat{A}_\ell-\frac{1}{2\ell+1}\sum_m\Gamma^{-1}_{\ell m \ell m} ,               
	\end{align}
where the left side of Eq. \ref{eq:Al2} is written as $\hat{A}'_\ell$ to denote an unbiased estimator. Therefore $\langle \hat{A}'_\ell \rangle = A_\ell$.

Note again that we relied on the fact that $\langle\hat{a}_{\ell m}\rangle \neq 0$. Namely, we are assuming $\langle\hat{a}_{\ell m}\rangle = a_{\ell m}$, where $a_{\ell m}$ is the true value of the clean map element. Therefore $\langle\hat{a}_{\ell m}\rangle$ is not an average over all possible realizations of the sky map, but rather an average over observations of our particular realization of the sky. In the former case, we would find $\langle\hat{a}_{\ell m}\rangle = 0$, and in the latter case, $\langle\hat{a}_{\ell m}\rangle = a_{\ell m}$. 

We can repeat the above procedure in the dirty map space. By again assuming $\langle \hat{x}_{\ell m}\rangle = x_{\ell m}\neq 0$, where $x_{\ell m}$ is the true value of the dirty map element, we'd find for the angular power spectrum:
	\begin{align}\label{eq:Xl2}
	  \hat{X}'_\ell  =  \hat{X}_\ell-\frac{1}{2\ell+1}\sum_m\Gamma_{\ell m \ell m},              
	\end{align}
where 

	\begin{align} \label{eq:xhat}
	 \hat{X}_\ell = \frac{1}{2\ell+1}\sum_m |\hat{x}_{\ell m}|^2               
	\end{align}
and, equivalently to Eq. \ref{eq:trueAl},
	\begin{align} \label{eq:xtrue}
	 {X}_\ell = \frac{1}{2\ell+1}\sum_m \langle|\hat{x}_{\ell m}|\rangle^2  .             
	\end{align}
    
\section{Dirty space SGWB auto-power angular power spectrum model}
\label{ap:model_auto}

Now let us consider the case in which we have a GW sky map resulting from a theoretical model $M$. This map may be in the dirty space, $x^M_{\ell m}$, or clean space, $a^M_{\ell m}$. 

For a given element of the clean space model map $a^M_{\ell m}$, we draw the real and imaginary components from the same Gaussian distribution $\mathcal{N}(0,\frac{1}{2}A^M_\ell)$ (or $\mathcal{N}(0,A^M_\ell)$ when $m=0$). The variance of a given map element is therefore

	\begin{align}\label{eq:varM}
	 var(  a^M_{\ell m}) = A^M_\ell = \langle |{a}^M_{\ell m}|^2 \rangle - \langle {a}^M_{\ell m}\rangle^2         
	\end{align}

    If we follow the same procedure as before, we obtain

    	\begin{align}\label{eq:AlM1}
	 \langle {A}^{RM}_\ell \rangle= \frac{1}{2\ell+1}\sum_m var(a^M_{\ell m}) + \frac{1}{2\ell+1}\sum_m\langle{a}^M_{\ell m}\rangle^2.               
	\end{align}

Here we use $A^{RM}_\ell$ to denote the model angular power spectrum reconstructed from its constituent $a^M_{\ell m}$ values. The distribution from which we are drawing the map elements explicitly has a mean of zero, so the expectation value in this context is referring to an average over realizations of the universe rather than an average over measurements of a single realization. Therefore we impose $\langle{a}^M_{\ell m}\rangle = 0$, leaving us with
    \begin{align}\label{eq:AlM2}
	   \langle {A}^{RM}_\ell \rangle= \frac{1}{2\ell+1}\sum_m var(a^M_{\ell m}) = \frac{1}{2\ell+1}\sum_m A^M_\ell = A^M_\ell,               
	\end{align}
\begin{align}
    \langle {A}^{RM}_\ell \rangle=A^M_\ell
\end{align}
which is the expected result. That is, if we draw many clean maps from a single angular power spectrum model, on average they should reconstruct the same values of $A^M_\ell$ from which they were drawn.

We now apply this process to the dirty map space. First, note how we convert to the dirty map space:
    \begin{align}\label{eq:convert}
	   x^M_{\ell m} = \Gamma_{\ell m, \ell' m'}a^M_{\ell'm'}            
	\end{align}
The variance of each dirty map element would therefore be
	\begin{align}
	 var(  x^M_{\ell m}) =  \langle |{x}^M_{\ell m}|^2 \rangle - \langle {x}^M_{\ell m}\rangle^2 = var(\Gamma_{\ell m, \ell' m'}a^M_{\ell' m'})        
	\end{align}
where
\begin{widetext}
	\begin{align}\label{eq:varxlm}
    var(\Gamma_{\ell m, \ell' m'}a^M_{\ell' m'})&= var \left(\sum_{\ell'=0}^{\ell_{max}}\sum_{m'=-\ell'}^{\ell'}\Gamma_{\ell m, \ell' m'}a^M_{\ell' m'} \right) \\ &= \sum_{\ell',m'}var(\Gamma_{\ell m, \ell' m'}a^M_{\ell' m'})+   2\sum_{\ell'm'< \ell''m''}Cov(\Gamma_{\ell m, \ell' m'}a^M_{\ell' m'},\Gamma_{\ell m, \ell'' m''}a^M_{\ell'' m''})\\
    &= \sum_{\ell',m'}var(\Gamma_{\ell m, \ell' m'}a^M_{\ell' m'}) \\
    &= \sum_{\ell',m'}|\Gamma_{\ell m, \ell' m'}|^2 var(a^M_{\ell' m'}) \\
    &= \sum_{\ell',m'}|\Gamma_{\ell m, \ell' m'}|^2 A^M_{\ell'}.
	\end{align}
\end{widetext}
Note that we assumed the covariance between different elements of the clean map is zero. We can then use the elements $x^M_{\ell m}$ to construct the angular power spectrum in the dirty map space:
\begin{align}\label{eq:XlM}
	 {X}^{M}_\ell = \frac{1}{2\ell+1}\sum_m|{x}^M_{\ell m}|^2.               
	\end{align}
We then arrive at a result analogous to Eq. \ref{eq:AlM1} by obtaining many sets of $x^M_{\ell m}$ and using each set to construct $X^M_\ell$:
\begin{align}\label{eq:XlM1}
	 \langle {X}^M_\ell \rangle= \frac{1}{2\ell+1}\sum_m var(x^M_{\ell m}) + \frac{1}{2\ell+1}\sum_m\langle{x}^M_{\ell m}\rangle^2.               
	\end{align}
Note again that this expectation is taken over many realizations of the drawn map elements $a^M_{\ell m}$ converted into the dirty map space. Since $\langle{a}^M_{\ell m}\rangle = 0$, we may also say that $\langle{x}^M_{\ell m}\rangle = 0$, so the second term on the right side of Eq. \ref{eq:XlM1} goes to zero, and we may write:

\begin{align}\label{eq:XlM2}
	 \langle {X}^M_\ell \rangle&= \frac{1}{2\ell+1}\sum_m var(x^M_{\ell m})\\
     &=\frac{1}{2\ell+1}\sum_m \sum_{\ell',m'}|\Gamma_{\ell m, \ell' m'}|^2 A^M_{\ell'}\\
     &=\frac{1}{2\ell+1}\sum_{m=-\ell}^\ell\left(  \sum_{\ell'=0}^{\ell_{max}}A^M_{\ell'} \left(\sum_{m'=-\ell'}^{\ell'}|\Gamma_{\ell m, \ell' m'}|^2 \right) \right)
	\end{align}

One can use this result to convert a clean space angular power spectrum model directly into the dirty space.

\section{Adding SGWB auto-power angular power spectrum model to detector noise in dirty space}
\label{ap:inject_auto}
Now if we add signal from a model $a^M_{\ell m}$ into noise-only data $\hat{a}_{\ell m}$, we have

\begin{align}\label{eqn:alm_inj}
\hat{a}^{sim}_{\ell m} = \hat{a}_{\ell m} + a^M_{\ell m}
\end{align}

\begin{align}
var(\hat{a}^{sim}_{\ell m}) = var(\hat{a}_{\ell m}) + var(a^M_{\ell m}) + 2Cov(\hat{a}_{\ell m},a^M_{\ell m}).
\end{align}

We assume the noise and signal are independent of each other, so the covariance term drops out and we are left with

\begin{align}
var(\hat{a}^{sim}_{\ell m}) =   \langle |\hat{a}^{sim}_{\ell m}|^2 \rangle - \langle \hat{a}^{sim}_{\ell m}\rangle^2=\Gamma^{-1}_{\ell m \ell m} + A^M_{\ell}
\end{align}

where we have substituted using Eq. \ref{eq:Exp4} and \ref{eq:varM}.

Now define the quantity
	\begin{align}
	 \hat{A}^{sim}_\ell = \frac{1}{2\ell+1}\sum_m |\hat{a}^{sim}_{\ell m}|^2.               
	\end{align}
The expected value can be written as
	\begin{align}
	 \langle \hat{A}^{sim}_\ell \rangle= \frac{1}{2\ell+1}\sum_m \langle|\hat{a}^{sim}_{\ell m}|^2\rangle. 
\end{align}
     	\begin{align}
	 \langle \hat{A}^{sim}_\ell \rangle= \frac{1}{2\ell+1}\sum_m var(\hat{a}^{sim}_{\ell m})+ \langle \hat{a}^{sim}_{\ell m} \rangle^2 
	\end{align}
Since this estimator is meant to estimate the true angular power spectrum, and not just its realization in our universe, we can assume $\langle{a}^M _{\ell m}\rangle = 0$. Further, since $\langle \hat{a} _{\ell m}\rangle = 0$ (averaged over detector noise realizations), we have $\langle \hat{a}^{sim}_{\ell m} \rangle = 0$, allowing us to write 
	\begin{align}
	 \langle \hat{A}^{sim}_\ell \rangle= \frac{1}{2\ell+1}\sum_m \left( \Gamma^{-1}_{\ell m \ell m} + A_\ell^M  \right)
\end{align}

	\begin{align}
	 = A^M_\ell+\frac{1}{2\ell+1}\sum_m  \Gamma^{-1}_{\ell m \ell m}.
\end{align}

We can then have unbiased estimators of $A^{sim}_\ell$ if we rearrange:


\begin{align}\label{eq:unbiased_inj2}
	 (\hat{A}^{sim}_\ell)'  =  A_\ell^M-\frac{1}{2\ell+1}\sum_m\Gamma^{-1}_{\ell m \ell m} .               
	\end{align}

    Now let's consider the equivalent scenario in the dirty-map space. The elements of the map with signal injection are given as
\begin{align}
\hat{x}^{sim}_{\ell m} = \hat{x}_{\ell m} + x^M_{\ell m}
\end{align}

We also define the variance of these elements as
\begin{widetext}
\begin{equation}
\begin{aligned}
var(\hat{x}^{sim}_{\ell m}) &= var(\hat{x}_{\ell m}) + var(x^M_{\ell m}) + 2Cov(\hat{x}_{\ell m},x^M_{\ell m}) \\
& = \Gamma_{\ell m, \ell m} + \sum_{\ell',m'}|\Gamma_{\ell m, \ell' m'}|^2 A^M_{\ell'} + 2Cov(\hat{x}_{\ell m}, x_{\ell m}^M) \\
& =\Gamma_{\ell m, \ell m} + \sum_{\ell',m'}|\Gamma_{\ell m, \ell' m'}|^2 A^M_{\ell'} + 2Cov(\hat{x}_{\ell m}, \left(\sum_{\ell'=0}^{\ell_{max}}\sum_{m'=-\ell'}^{\ell'}\Gamma_{\ell m, \ell' m'}a^M_{\ell' m'} \right)) \\
& =\Gamma_{\ell m, \ell m} + \sum_{\ell',m'}|\Gamma_{\ell m, \ell' m'}|^2 A^M_{\ell'} + 2\sum_{\ell'=0}^{\ell_{max}}\sum_{m'=-\ell'}^{\ell'}Cov(\hat{x}_{\ell m}, \Gamma_{\ell m, \ell' m'}a^M_{\ell' m'} ) \\
& =\Gamma_{\ell m, \ell m} + \sum_{\ell',m'}|\Gamma_{\ell m, \ell' m'}|^2 A^M_{\ell'} + 2\sum_{\ell'=0}^{\ell_{max}}\sum_{m'=-\ell'}^{\ell'}\Gamma_{\ell m, \ell' m'}Cov(\hat{x}_{\ell m}, a^M_{\ell' m'} ) \\
& =\Gamma_{\ell m, \ell m} + \sum_{\ell',m'}|\Gamma_{\ell m, \ell' m'}|^2 A^M_{\ell'} .
\end{aligned}
\end{equation}
\end{widetext}
Estimators of the angular power spectrum are defined as
	\begin{align} \label{eq:Xinj_est}
	 \hat{X}^{sim}_\ell = \frac{1}{2\ell+1}\sum_m |\hat{x}^{sim}_{\ell m}|^2.               
	\end{align}
with expectation value
\begin{equation}
\begin{aligned}
	\langle \hat{X}^{sim}_\ell \rangle & = \frac{1}{2\ell+1}\sum_m \langle|\hat{x}^{sim}_{\ell m}|^2\rangle \\
	 &=\frac{1}{2\ell+1}\sum_m var(\hat{x}^{sim}_{\ell m})+ \langle \hat{x}^{sim}_{\ell m} \rangle^2 .   
\end{aligned}
\end{equation}
Substituting from Eqs. \ref{eq:varxlm} and \ref{eq:Exp1} we may rewrite as

\vspace{1em}
\begin{widetext}
\begin{equation}
\begin{aligned} \label{eq:xinj48}
	\langle \hat{X}^{sim}_\ell \rangle & = \frac{1}{2\ell+1}\sum_m \left( \Gamma_{\ell m ,\ell m} + \sum_{\ell',m'}|\Gamma_{\ell m, \ell' m'}|^2 A^M_{\ell'} + \langle \hat{x}_{\ell m} + x^M_{\ell m}\rangle^2\right) \\
    &=  \frac{1}{2\ell+1}\sum_m \left((\langle \hat{x}_{\ell m}\rangle + \langle x^M_{\ell m} \rangle)^2+\Gamma_{\ell m ,\ell m} + \sum_{\ell' = 0}^{\ell_{max}}A_{\ell'}^M \sum_{m'=-\ell'}^{\ell'}|\Gamma_{\ell m, \ell' m'}|^2 \right) \\
    &=\frac{1}{2\ell+1}\sum_m \left(\Gamma_{\ell m ,\ell m} + \sum_{\ell' = 0}^{\ell_{max}}A_{\ell'}^M \sum_{m'=-\ell'}^{\ell'}|\Gamma_{\ell m, \ell' m'}|^2 \right) \\
    &=  \langle X^M_\ell \rangle + \frac{1}{2\ell+1}\sum_m \Gamma_{\ell m ,\ell m}
\end{aligned}
\end{equation}
\end{widetext}
In the second line of Eq. \ref{eq:xinj48} we assume $\langle x^M_{\ell m} \rangle=0$ (again, averaging over many sky realizations), and in the fourth line we use the result from Eq. \ref{eq:XlM2}. Our biased estimator can therefore be written as

\begin{equation}
\begin{aligned} \label{eq:xinj2}
	\langle \hat{X}^{sim}_\ell \rangle & \approx \langle X^M_\ell \rangle +\frac{1}{2\ell+1}\sum_m \left( \Gamma_{\ell m ,\ell m}\right) \\,
\end{aligned}
\end{equation}
allowing us to write an expression for the unbiased estimators 
\begin{equation}
\begin{aligned} \label{eq:xinj_unbiased}
	(\hat{X}^{sim}_\ell)' & = \hat{X}_\ell^{sim} -\frac{1}{2\ell+1}\sum_m \left( \Gamma_{\ell m ,\ell m}\right) \\,
\end{aligned}
\end{equation}

where $\hat{X}_\ell^{sim}$ is given in Eq. \ref{eq:Xinj_est}.

\section{Dirty space SGWB-EM angular cross-power spectrum model}\label{app:deriv-of-analytical-model-Zl}
Let us consider the case in which we have a GW sky map and EM tracer sky map resulting from a theoretical model M. This map may be in the dirty space, $x_{\ell m}^M$ and $y_{\ell m}^M$, or the clean space, $a_{\ell m}^M$ and $b_{\ell m}^M$, for the GW map and EM tracer maps respectively. We begin in the clean space.

For given elements of the clean space maps $a_{\ell m}^M$ and $b_{\ell m^{'}}^M$, we draw the real and imaginary components from a bi-variate Gaussian distribution $\mathcal{N}(\boldsymbol{0},\Sigma_{ab})$ where $\Sigma_{ab}$ is given by

\begin{equation}
    \Sigma_{ab} = \begin{pmatrix}
        \frac{1}{2}A_\ell^M & \frac{1}{2}C_\ell^M \\
        \frac{1}{2}C_\ell^M & \frac{1}{2}B_\ell^M
    \end{pmatrix}
\end{equation}
(or $2\Sigma_{ab}$ when $m=0$).
The covariance of the given map elements is therefore
\begin{equation}
    \mathrm{Cov}(a_{\ell m}^M,b_{\ell m}^M) = C_\ell^M = \langle (a_{\ell m}^M)^* b_{\ell m}^M \rangle - \langle (a_{\ell m}^M)^* \rangle\langle b_{\ell m}^M \rangle
\end{equation}

The model clean cross-power spectrum is given by

\begin{equation}
    C^M_\ell = \frac{1}{2\ell+1} \sum_{m=-\ell}^\ell \left( a_{\ell m}^M\right)^* b_{\ell m}^M
\end{equation}
Find the expectation value of both sides:
\begin{align}
    \langle C_\ell^M\rangle &= \frac{1}{2\ell+1} \sum_{m} \langle \left(a_{\ell m}^M \right)^* b_{\ell m}^M\rangle & \\
    &= \frac{1}{2\ell+1} \sum_{m} \big[\langle (a_{\ell m}^M)^* \rangle\langle b_{\ell m}^M \rangle + \\ 
    & \qquad \qquad \qquad \qquad \quad  + \mathrm{Cov}(a_{\ell m}^M,b_{\ell m}^M)  \big] \nonumber
\end{align}

We are assuming that $\langle a_{\ell m}^M \rangle = \langle b_{\ell m}^M \rangle = 0 $ as we are averaging over all possible realizations of the universe. This leaves us with 

\begin{align}
    \langle C_\ell^M\rangle &= \frac{1}{2\ell+1} \sum_{m}  \mathrm{Cov}\left(a_{\ell m}^M,b_{\ell m}^M\right) \\
    &= \frac{1}{2\ell+1} \sum_{m} C_\ell^M \\ 
    &= \frac{C_\ell^M}{2\ell+1} (2\ell+1) \\
    \langle C_\ell^M\rangle &= C_\ell^M
\end{align}
which is the expected result. That is, if we draw many clean maps from the power spectra models, on average they should reconstruct the same values of $C_\ell^M$ from which they are drawn. 

Applying this process to the dirty map space, we first note how to convert to the dirty space for EM tracers:
\begin{equation}\label{eqn:dirty-blms_mc}
    y_{\ell m}^M = \sum_{\ell'm'}\Gamma_{\ell m,\ell ' m'} b_{\ell ' m'}^M
\end{equation}
The covariance of given map elements would therefore be
\begin{equation}
    \mathrm{Cov}\left(x_{\ell m}^M,y_{\ell m}^M\right) = \langle (x_{\ell m}^M)^* y_{\ell m}^M \rangle - \langle (x_{\ell m}^M)^* \rangle\langle y_{\ell m}^M \rangle
\end{equation}
where
\begin{widetext}
    \begin{align}
    \mathrm{Cov}(x_{\ell m}^M,y_{\ell m}^M) &= \mathrm{Cov}\left(\sum_{\ell'm'}\Gamma_{\ell m,\ell ' m'} a_{\ell' m'}^M,\sum_{\ell''m''}\Gamma_{\ell m,\ell '' m''}b_{\ell'' m^{''}}^M\right) \\
    &= \sum_{\ell'm'}\sum_{\ell''m''} \mathrm{Cov} \left( \Gamma_{\ell m,\ell ' m'} a_{\ell' m'}^M, \Gamma_{\ell m,\ell '' m''}b_{\ell'' m^{''}}^M  \right)  \\
    &= \sum_{\ell'm'}\sum_{\ell''m''} \Gamma_{\ell m,\ell ' m'} \mathrm{Cov} \left( a_{\ell' m'}^M, b_{\ell'' m^{''}}^M \right) \Gamma_{\ell m,\ell '' m''}^T
\end{align}
\end{widetext}

and $\mathrm{Cov} \left( a_{\ell' m'}^M, b_{\ell'' m^{''}}^M \right)$ is only non-zero when $\ell' = \ell''$ and $m'=m''$  and $\Gamma^T_{\ell m,\ell''m''} = \Gamma_{\ell''m'', \ell m}$ so
\begin{align}
    \mathrm{Cov}(x_{\ell m}^M,y_{\ell m}^M) &= \sum_{\ell',m'} \Gamma_{\ell m,\ell ' m'} \mathrm{Cov} \left( a_{\ell' m'}^M, b_{\ell' m^{'}}^M \right) \Gamma_{\ell'm', \ell m} \\
    &= \sum_{\ell',m'} \Gamma_{\ell m,\ell ' m'} C_{\ell'}^M \Gamma_{\ell'm', \ell m}
\end{align}

We can write the SGWB-EM tracer cross-power in the dirty space:
\begin{align}\label{eqn:Zl-alm-blm}
    Z_\ell^M &= \frac{1}{2\ell+1} \sum_m \left( x_{\ell m}^M \right)^* y_{\ell m}^M
\end{align}

and following the same process as used before, we can arrive at an equation for the expectation value of $Z_\ell^M$:

\begin{align}\label{eqn:expZl}
    \langle Z_\ell^M \rangle &= \frac{1}{2\ell+1} \sum_m  \mathrm{Cov}(x_{\ell m}^M,y_{\ell m}^M) + \\
    & \qquad+\frac{1}{2\ell+1}  \langle (x_{\ell m}^M)^* \rangle\langle y_{\ell m}^M \rangle \nonumber
\end{align}
Since $\langle a_{\ell m}^M \rangle = 0$ and $\langle b_{\ell m}^M \rangle=0$, we may also say that $\langle x_{\ell m}^M \rangle = 0$ and $\langle y_{\ell m}^M \rangle=0$, so the second term of Eq. \ref{eqn:expZl} goes to zero and we may write

\begin{align}
    \langle Z_\ell^M \rangle &= \frac{1}{2\ell+1} \sum_m \mathrm{Cov}(x_{\ell m}^M,y_{\ell m}^M) \\
    &= \frac{1}{2\ell+1} \sum_{m=-\ell}^\ell  \sum_{\ell'=0}^{\ell_{max}}  \sum_{m'=-\ell'}^{\ell'} \Gamma_{\ell m,\ell ' m'}  C_{\ell'}^M  \Gamma_{\ell'm', \ell m} 
    \label{Eq:ZlM}
\end{align}

We used this result to convert a clean space angular cross-power spectrum model directly into the dirty space.

\section{Adding SGWB-EM angular cross-power spectrum model to detector noise in dirty space}
Now consider the case where we are injecting a signal from a model into noise-only GW data and assume that the EM tracer data has no associated noise. We have $\hat{a}_{\ell m}^{sim}$, given by Eq. \ref{eqn:alm_inj}, and $b_{\ell m}^M$ (since we are assuming that $b_{\ell m}$'s have no detector noise), yielding

\begin{align}
    \mathrm{Cov}(\hat{a}_{\ell m}^{sim},b_{\ell m}^M) &= \langle (\hat{a}_{\ell m}^{sim})^* b_{\ell m}^M \rangle - \langle (\hat{a}_{\ell m}^{sim})^* \rangle\langle b_{\ell m}^M \rangle \\
    & = \mathrm{Cov}(\hat{a}_{\ell m},b_{\ell m}^M) + \mathrm{Cov}(a_{\ell m}^{M},b_{\ell m}^M)
\end{align}
We assume the noise and signal are independent of each other, so $\mathrm{Cov}(\hat{a}_{\ell m},b_{\ell m}^M)=0$ and $\langle b_{\ell m}^M \rangle=0$. We are left with
\begin{align}
    \mathrm{Cov}(\hat{a}_{\ell m}^{sim},b_{\ell m}^M) &= \langle (\hat{a}_{\ell m}^{sim})^* b_{\ell m}^M \rangle \\ 
    &=  \mathrm{Cov}(a_{\ell m}^{M},b_{\ell m}^M) \\
    &= C_\ell^M 
\end{align}
Now we have the clean space signal injected cross-power, 
\begin{equation}
    \hat{C}_\ell^{sim} = \frac{1}{2\ell+1} \sum_m \left( \hat{a}_{\ell m}^{sim}\right)^* b_{\ell m}^M
\end{equation}
and the expected value can be written as 
\begin{align}
    \langle \hat{C}_\ell^{sim} \rangle &= \frac{1}{2\ell+1} \sum_m \langle (\hat{a}_{\ell m}^{sim})^* b_{\ell m}^M \rangle \\
    &= \frac{1}{2\ell+1} \sum_m \mathrm{Cov}(\hat{a}_{\ell m}^{sim},b_{\ell m}^M) \\
    &= \frac{1}{2\ell+1} \sum_m \mathrm{Cov}(a_{\ell m}^{M},b_{\ell m}^M) \\
    &= \frac{1}{2\ell+1} \sum_m C_\ell^M \\
    \langle \hat{C}_\ell^{sim} \rangle &= C_\ell^M
\end{align}
meeting our expectations.

Now, if we consider an equivalent scenario in the dirty space, the elements of the signal injected dirty space map, $x_{\ell m}^{sim}$ and $y_{\ell m}^M$ have covariance given by
\begin{align}
    \mathrm{Cov}\left(\hat{x}_{\ell m}^{sim},y_{\ell m}^M\right) &= \mathrm{Cov}\left(x_{\ell m}^{M},y_{\ell m}^M\right) \\
    &= \sum_{\ell',m'} \Gamma_{\ell m,\ell ' m'} C_{\ell'}^M \Gamma_{\ell'm', \ell m} \\
    &= \langle (\hat{x}_{\ell m}^{sim})^* y_{\ell m}^M \rangle
\end{align}
The injected cross-power in the dirty space is defined as
\begin{equation}
    \hat{Z}_\ell^{sim} = \frac{1}{2\ell+1}\sum_m \left( \hat{x}_{\ell m}^{sim}\right)^* y_{\ell m}^M 
\end{equation}
with expectation value
\begin{align}
    \langle \hat{Z}_\ell^{sim} \rangle &= \frac{1}{2\ell+1} \langle (\hat{x}_{\ell m}^{sim})^* y_{\ell m}^M \rangle \\
    &= \frac{1}{2\ell+1} \sum_m \mathrm{Cov}(\hat{x}_{\ell m}^{sim},y_{\ell m}^M) \\
    &= \frac{1}{2\ell+1} \sum_m \mathrm{Cov}(x_{\ell m}^{M},y_{\ell m}^M) \\
    &= \frac{1}{2\ell+1} \sum_m \sum_{\ell',m'} \Gamma_{\ell m,\ell ' m'} C_{\ell'}^M \Gamma_{\ell'm', \ell m} \\
    \langle \hat{Z}_\ell^{sim} \rangle &= \frac{1}{2\ell+1} \sum_{m=-\ell}^\ell   \sum_{\ell'=0}^{\ell_{max}}  \sum_{m'=-\ell'}^{\ell'} \Gamma_{\ell m,\ell ' m'}  C_{\ell'}^M  \Gamma_{\ell'm', \ell m}
\end{align}
which, again, is similar to the expression that we get for the model in Eq. \ref{Eq:ZlM}. With enough data to average over, we expect to be able to recover the signal generated by the model, so this result is as expected, again without the bias correction term you see in the auto-power case.

\subsection{Estimating $K_{draw}^Z$}\label{app:crossspower-Kdraw}

When injecting a signal into noise, similar to the auto-power case, we introduce additional uncertainty coming from the draw of the model map elements from the multivariate Gaussian distribution (Eq. \ref{eqn:Sigma_ab}). We estimate this uncertainty by drawing the map elements in the clean space, transforming to the dirty space (Eq. \ref{eqn:dirtying_alm_mc} and \ref{eqn:dirty-blms_mc}), and calculating the model cross-power (Eq. \ref{eqn:Zl-alm-blm}). We 
repeat this process $10^4$ times to calculate the covariance matrix $K_{draw}^Z$ with elements given by
\begin{equation}
    \left(K_{draw}^Z\right)_{\ell,\ell'} = \langle Z_\ell^M Z_{\ell'}^M \rangle - \langle Z_\ell^M \rangle \langle Z_{\ell'}^M \rangle.
\end{equation}

\bibliography{refs}
\bibliographystyle{apsrev4-1}

\end{document}